\documentclass[useAMS, usenatbib]{mn2e} 

\usepackage{graphicx}

\title{Cosmic ray acceleration in young supernova remnants}
\author[K.M. Schure \& A.R. Bell]{K.M. Schure$^{1}$\thanks{E-mail: K.Schure1@physics.ox.ac.uk} and A.R. Bell$^{1}$\\
$^{1}$Department of Physics, University of Oxford, Clarendon Laboratory, Parks Road, Oxford OX1 3PU, United Kingdom}

\begin{document}

\newcommand\araa{{ARA\&A}}
\newcommand\apj{{ApJ}}
\newcommand{\apjl}{ApJL}
\newcommand\apjs{{ApJS}}
\newcommand\aap{{A\&A}}
\newcommand\mnras{{MNRAS}}
\newcommand\rmxaa{{Rev. Mexicana Astron. Astrofis.}}
\newcommand\nat{{Nature}}
\newcommand\physrep{{Phys.~Rep.}}
\newcommand\memsai{{MmSAI}}
\newcommand\ssr{{Space Science Rev.}}

\date{\ldots; \ldots}
\pagerange{\pageref{firstpage}--\pageref{lastpage}}\pubyear{2010}
\maketitle
\label{firstpage}

\begin{abstract}
We investigate the appearance of magnetic field amplification resulting from a cosmic ray escape current in the context of supernova remnant shock waves. The current is inversely proportional to the maximum energy of cosmic rays, and is a strong function of the shock velocity. Depending on the evolution of the shock wave, which is drastically different for different circumstellar environments, the maximum energy of cosmic rays as required to generate enough current to trigger the non-resonant hybrid instability that confines the cosmic rays follows a different evolution and reaches different values. We find that the best candidates to accelerate cosmic rays to $\sim$ few PeV energies are young remnants in a dense environment, such as a red supergiant wind, as may be applicable to Cassiopeia A. We also find that for a typical background magnetic field strength of $5\;\mu$G the instability is quenched in about 1000 years, making SN1006 just at the border of candidates for cosmic ray acceleration to high energies. 
\end{abstract}

\begin{keywords}
MHD --- ISM: cosmic rays --- ISM: supernova remnants --- acceleration of particles --- instabilities
\end{keywords}

\section{Introduction}
\label{sec:intro}

Cosmic rays were discovered around 100 years ago by Victor Hess. He showed with a balloon experiment that the ionising radiation increases with altitude and therefore has a cosmic origin. The misleading term `cosmic rays' was coined before it was discovered that the radiation consisted of particles rather than electromagnetic waves. It is now known that they are ionised particles that travel near the speed of light, and can be anything from electrons, protons, to ionised higher-Z elements. 
Cosmic rays with energies to the `knee', being the break in the spectrum at around 3 PeV ($10^{15}$~eV), are believed to be of Galactic origin. More energetic ones are not effectively confined by the Galactic magnetic field.

Over the past decades various theories have been designed as to explain the origin of Galactic cosmic rays. The preferred one is the acceleration of cosmic rays in the process of `diffusive shock acceleration' (DSA) in supernova remnants (SNRs) \citep[see e.g.~][for a review]{2001MalkovDrury}. Various aspects of the observations can be nicely explained: the energy budget in SNRs is enough to re-energise the cosmic ray population, the acceleration is quick enough to accelerate cosmic rays to around PeV energies if the magnetic field is amplified, and the magnetic field can be amplified by the cosmic rays themselves when a fraction of the cosmic rays escape to trigger an amplifying instability, of which the fastest growing one on relevant scales is the non-resonant hybrid (NRH) instability \citep{2000LucekBell, 2004Bell, 2005Bell}. 

In this paper we aim to connect the theory of magnetic field amplification by the NRH instability to the confinement of cosmic rays in the vicinity of the SNR blast wave and to evaluate the maximum energy for the cosmic rays. The growth rate of the instability and the confinement are non-linearly intertwined, as more efficient confinement reduces the upstream growth rate, which in turn reduces confinement, and vice versa in a self-regulating manner. Similar types of studies have been performed by various authors. For example, \citet{2008Zirakashvilietal} numerically modelled the instability in plane parallel geometry, and  looked in more detail at the downstream region in \citet{2008ZirakashviliPtuskin}. The same model was used to calculate the Galactic cosmic ray spectrum in \cite{2010Ptuskinetal}. \citet{2009Revilleetal} used a similar calculation of the NRH instability to determine the free-escape boundary, a popular concept in calculations and simulations of cosmic rays in non-linear steady-state models, such as used by e.g. \citet{1983Eichler,1984EllisonEichler,2009Kangetal,2009Vladimirovetal,2010Capriolietal,2010Ohiraetal,2010Patnaudeetal, 2012Kang}. We aim to highlight the connection between the instability and its effect on the maximum cosmic ray energy, and how they can be reliably calculated and modelled in a spherical geometry such that the numerical results can be interpreted in a quantitative sense. 

A first-principle simulation of the NRH instability has recently been done by \citet{2013Belletal}, where a combination of MHD with a VFP description of the cosmic rays to second order anisotropy in momentum has been shown to work: a population of escaping cosmic rays triggers the instability that generates enough magnetic turbulence to confine more recent cosmic rays to the shock region. Details of how this affects the diffusion approach that should be assumed for cosmic rays are still unknown, but observational evidence hints that it is close to Bohm in the sources that we think produce the high-energy end of the Galactic cosmic rays \citep{2006Stageetal, 2007Uchiyamaetal}. The simplification in this work, of approaching the cosmic rays as a source current in the MHD code, allows for simulations on larger scales such as applicable in SNR shock waves.

In Section~\ref{sec:currentsource} we describe the link between the magnetic field growth and the maximum energy of the accelerated cosmic rays. In Section~\ref{sec:model} we quantitatively show the various stages of the magnetic field amplification as it results from the NRH instability in MHD simulations. In Section~\ref{sec:Emax} we describe how the maximum cosmic ray energy can be self-consistently calculated as a function of time, requiring enough escaping cosmic rays to let the NRH instability grow fast enough to confine the cosmic rays. We assess the results in light of the various types of SNe and environments. In Section~\ref{sec:cumulative} we evaluate what this means in terms of which sources can accelerate PeV cosmic rays, and how the various sources add up to make up the Galactic CR population. In Section~\ref{sec:SNRmodels} we illustrate the result of this analysis with some simulations, where we integrate the previously calcuated cosmic-ray current with simulations of the evolution of young SNRs. 
Finally, we reiterate the implications of our results in Section~\ref{sec:discussion}.

\section{Cosmic ray current source}
\label{sec:currentsource}
If cosmic rays are efficiently accelerated at the shock, it can be assumed that a certain fraction $\chi$ of the shock's kinetic energy is transferred to the relativistic particle population and that a fixed fraction of this energy is carried upstream by escaping CR.
For a given CR energy flux the electric current carried by the CR is inversely
proportional to the energy of each escaping CR.
The return current that is consequently carried by the thermal plasma to maintain neutrality can give rise to a number of current driven instabilities \citep[for an overview see][]{2012Schureetal}.

We are investigating the effect of this current on the magnetic field growth and stability of the upstream plasma in the regime of the non-resonant hybrid instability \citep{2004Bell}. To our knowledge, this is the instability that allows for the fastest magnetic field growth in plasmas representative of young SNRs. Its maximum growth rate $\gamma$ and corresponding wave number $k$ are (in cgs units): 
\begin{eqnarray}
\label{eq:gammamax}
\gamma_{max}=\frac{j}{c}\sqrt{\frac{\pi}{\rho}}, \qquad k_{max}=\frac{2 \pi j}{B c},
\end{eqnarray}
with $j$ the return current resulting from balancing the escape current of cosmic rays, $B$ the magnetic field parallel to $j$, and $\rho$ the density of the plasma.
If we, for now, assume that the instability grows at the maximum rate, the number of e-foldings for magnetic field amplification at a fixed radius $R$ upstream of (or at) the variable shock radius $R_s$ is given by:
\begin{eqnarray}
\label{eq:gt_int}
\gamma \tau = \int_0^R \frac{j(R_s)}{c}\sqrt{\frac{\pi}{\rho(R)}}\frac{d R_s}{u_s},
\end{eqnarray}
with $u_s$ the shock velocity.

A surface area dilution applies in the entire upstream region, resulting in a $(R_s/R)^2$ dependency for $j$ at a radius $R$ when the CR are accelerated by the shock at radius $R_s$.
Fixing the cosmic ray energy density $U_{cr}$ as a fraction of the shock kinetic energy, by fixing $\chi$ and defining $U_{cr} =\chi \rho u_s^2$ gives:
\begin{eqnarray}
\label{eq:j_pz}
j=\frac{\chi \rho (R_s) u_s^3 q}{E_{max} \ln(E_{max}/m_p c^2)} \left(\frac{R_s^2}{R^2}\right).
\end{eqnarray}
for the current from escaping cosmic rays with charge $q$ upstream of the shock, given the number of cosmic rays as a function of energy follows a powerlaw with a slope $dN/dE \propto E^{-2}$ (or $N \propto E^{-1}$) \citep[see also][]{2013Belletal}. There are various indications that the source spectrum in fact deviates from the canonical powerlaw slope as derived from diffusive shock acceleration. A flatter spectrum is expected if nonlinear shock modification is taken into account \citep[e.g.~][]{2006Vladimirovetal}, whereas oblique high-velocity shocks, or other reasons, may generate a steeper spectrum \citep[e.g.~][]{2011Belletal}. Clearly, the number of escaping cosmic rays and thus the current becomes a different function of $E_{max}$. In a related analysis applied to planar shocks propagating at constant velocity, \citet{2008ZirakashviliPtuskin} discuss how $\chi$ (their $\eta_{esc}$) varies in response to non-linear CR pressure feedback onto the shock structure and the resulting effect on $E_{max}$.  The crucial factor determining $E_{max}$ is the fraction of the available energy flux $\rho u_s^3$ given to CR at the highest energy.

Depending on whether $\rho$ now has a dependency on $R$ as in a stellar wind environment (circumstellar medium, CSM), or not, as in a standard Type Ia environment (interstellar medium, ISM), solving the integral equation \ref{eq:gt_int} for fixed $u_s$ constrains $E_{max}$ to behave as:
\begin{eqnarray}
\label{eq:emax}
E_{max}&=&\frac{\chi u_s^2 q R_s \sqrt{\rho \pi}}{c \gamma \tau \ln(E_{max}/(m_p c^2))} \qquad (\textsc{CSM})\\\nonumber
E_{max}&=&\frac{\chi u_s^2 q R_s \sqrt{\rho \pi}}{2 c \gamma \tau \ln(E_{max}/(m_p c^2))} \qquad (\textsc{ISM}).
\end{eqnarray}
This is the energy at which the particle spectrum starts to roll over. The difference between this description of $E_{max}$ and that constrained by either the Hillas criterion that restricts the energy based on the size of the system \citep{1984Hillas}, and the Lagage-Cesarsky requirement \citep{1983LagageCesarsky} that the rate at which energy is gained determines the maximum energy, is the inclusion and outcome of magnetic field amplification. Given the magnetic field is amplified to levels that are reasonably expected in the presence of the NRH instability, both Hillas and Lagage-Cesarsky limits are less restrictive on $E_{max}$ than the fact that an escape current is needed to amplify the field to levels that allow for confinement of the cosmic rays. In other words, even though $E_{max}$ is lower using the Hillas and Lagage-Cesarsky limits with the background magnetic field strength, allowing for magnetic field growth alleviates this restriction and, if enough amplification is required by using the appropriate value for $\gamma \tau$ as we will discuss below, $E_{max}$ subsequently is restricted by allowing for enough escape current rather than by time dependence or size of the accelerator.

By substituting Eq.~\ref{eq:emax} into Eq.~\ref{eq:j_pz}, and evaluating for $R=R_s$, we can write the current in terms of $\gamma \tau$ as:
\begin{eqnarray}
\label{eq:j_gt}
j(R_s)=\gamma \tau \frac{c u_s}{R_s}\sqrt{\frac{\rho}{\pi}} \quad (\textsc{CSM})\\\nonumber
j(R_s)=2 \gamma \tau \frac{c u_s}{R_s}\sqrt{\frac{\rho}{\pi}} \quad (\textsc{ISM}).
\end{eqnarray}

Which values of $\chi$ and $\gamma \tau$ should be used are up for discussion and should ultimately be determined by comparison with observations. From self-similar dimensional arguments we expect $\chi$ and $\gamma \tau$
to be independent of the shock velocity provided the Mach number is large
and $E_{max}$ is much greater than the proton rest mass energy, 
but this assumption can be questioned. However, to first approximation some bounds for these parameters can be estimated, where we assume they are independent of the absolute value for the shock velocity.
The parameter $\chi$ is a measure of how efficient energy is transferred to the cosmic ray population, as a fixed fraction of the downstream thermal energy density. From the Rankine-Hugoniot jump conditions, the downstream (post-shock) thermal energy density is $U_{th}=\frac{9}{8}\rho_0 u_s^2$. If 30\% of that is transferred to cosmic rays, this translates to a value for $\chi=0.34$ (to be compared to $\eta \ln(E_{max}/E_{rest}$) in \citet{2013Belletal} and others).

A rough lower and upper bound on $\gamma \tau$ can be estimated in various ways. A lower limit can be found by requiring $B_\perp \equiv  \sqrt{(B_2^2+B_3^2)/2} \approx B_0$ (where $B_0$ is the background field, aligned with the first dimension, and $B_\perp$ the magnitude of the magnetic field in the directions perpendicular to that) for confinement (necessary for Bohm diffusion), such that $\gamma \tau > \ln(B_0/B_{\perp,0})$.  If the initial fluctuation is around 5\% of that of the background magnetic field, this translates to a value of $\gamma \tau > 3$ as a very minimal value (and upper limit to $E_{max}$). Any value higher than that, will require a lower energy cut-off in the spectrum to allow for a higher escape current. A tentative upper bound can be estimated by looking at the value for the saturated magnetic field, and approximating that growth does not slow down, which in reality it will. As such, this yields a lower limit on the upper bound, since, as we will see in the next section, $\gamma$ will decrease when $\langle B_\perp\rangle /B_0 \approx 2-3$, where $\langle B_\perp \rangle$ is the average of the magnitude $B_\perp$ over the 2nd dimension (in 2D simulations). Very early on, while the cosmic ray energy is not yet built up, the $dE/dt$ restriction as posed in \citet{1983LagageCesarsky} causes $E_{max}$ to be lower than described by Eq.~\ref{eq:emax}. This results initially in a large escape current and thus a temporary very high value for $\gamma \tau$, allowing for more rapid growth of $E_{max}$. We find that within the first year, this allows for the cosmic ray energy to reach levels that from thereon are restricted by the magnetic field amplification requirement. At no point at these early stages does the magnetic field as required for sufficient amplification exceed the saturation value of the magnetic field, which we will discuss next.

The value for the saturated magnetic field can be estimated by looking at when $c/4 \pi (\nabla \times B)$ becomes comparable to the cosmic ray current, which is when the size of the fluctuations and the gyroradius of the cosmic rays are approximately equal \citep{2004Bell}, such that $r_g \approx \lambda\ (= 2 \pi/k)$, meaning $E/qB \approx Bc/2j$, with $j \approx ckB/(4\pi)$ the current at the smallest growing scale, being at $k=2k_{max}$.
This gives an estimate of the saturation magnetic field 
$\sqrt{{4 \pi \chi \rho u_s^3}/{c \ln(E_{max}/(m_p c^2))}}$
on the scale of the Larmor radius
of a particular CR.
This spatial component of the magnetic energy density is independent of the CR energy.
If we suppose that the magnetic energy density per logarithmic interval in scalelength
is the same on all
scales ranging from the Larmor radius of a GeV proton to the Larmor radius of a PeV proton,
the estimated characteristic saturation magnetic field integrated over all scales is
\begin{eqnarray}
\label{eq:Bsat}
B_{sat}\sim\sqrt{\frac{4 \pi \chi \rho u_s^3}{c}}.
\end{eqnarray}
For $B_{sat} \gg B_0$, and $B_{nl}^2\approx B_0^ 2(B_{\perp,0}/B_0)^2 \exp(2 \gamma \tau)$, we then find the following `upper limit' to $\gamma \tau$ of
\begin{eqnarray}
\label{eq:gammatsat}
\gamma \tau < \left(\frac{1}{2}\right) \ln \left(\frac{4 \pi \chi \rho u_s^3}{c B_0^2\left(B_{\perp,0} / B_0 \right)^2}\right).
\end{eqnarray}
Because of the logarithmic dependence on the initial fluctuations, $\gamma \tau$ is not very sensitive to which value we assume for that. Putting in numbers provides an upper bound of $\gamma \tau \approx 9$, a factor 3 larger than our absolute lower limit. We will get back to this with our simulations and numerical solutions in Sections~\ref{sec:Emax} and \ref{sec:SNRmodels}. It should be kept in mind that a factor of 3 in $\gamma \tau$ is only a fractional change in $E_{max}$ and the current, but an exponential change in the magnetic field.

\section{Current driven magnetic field amplification}
\label{sec:model}

We use the MHD code from the AMRVAC framework \citep{2007HolstKeppens} to simulate the NRH instability on a 2.5D grid (2 spatial coordinates, and vector components in all 3 directions). For illustration purposes of the instability in its purest form, in this section we will first show the results of simulations on a cartesian grid in the $x$-$y$ plane, with a fixed current, while later, in Sect.~\ref{sec:SNRmodels} we will run simulations in a spherical geometry. The current is implemented as an imposed source term, where in the momentum equation a term is added that represents the Lorentz force resulting from this current. By fixing the current in the direction of the first dimension ($x$ on a cartesian grid), we only have to add the Lorentz force in the form of $j \times B$ to the $y$ and $z$ components of the momentum equation. The added momentum adds to the total energy, resulting in a net energy gain throughout the simulation that in reality should be at the cost of the shock kinetic energy. However, the increase is very small compared to the total kinetic energy and does not change the dynamics.

We initialise the plasma with a homogeneous density, zero velocity, and a magnetic field that consists of a background field $B_0$ parallel to the current ($x$-direction) and a random component of around 5\% of that of the background field, and the length scale of the fluctuations is approximately the wavelength of the most unstable mode.

The instability has a maximum growth rate for a wavelength of $\lambda_{min}=2\pi/k_{max}$, with $k_{max}$ the wavenumber for which the growth rate starts to decrease, as given by Eq.~\ref{eq:gammamax}, such that
\begin{eqnarray}
\lambda_{min}\approx\frac{B_0 c}{j}.
\end{eqnarray}
In order to be able to resolve this maximum growth rate, the size of the grid cells needs to be smaller than $\lambda_{min}$ and the fluctuations seeded on the right scale.

\begin{figure}
  \centering
 \includegraphics[angle=-90,width=0.45\textwidth]{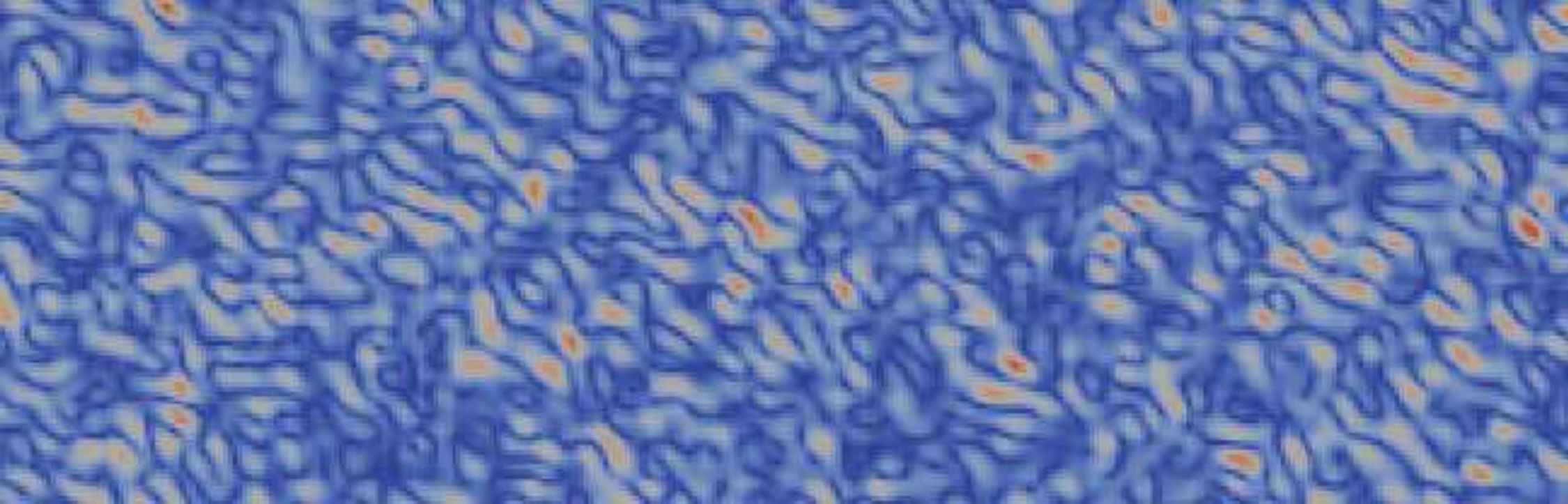}
 \includegraphics[angle=-90,width=0.45\textwidth]{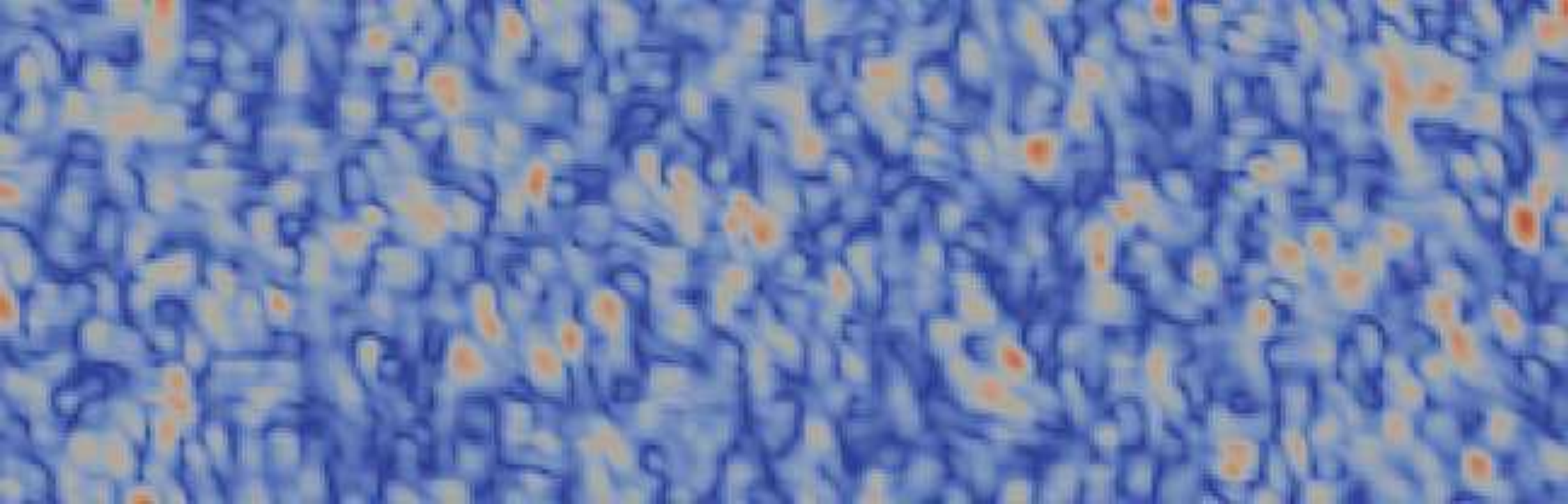}
 \includegraphics[angle=-90,width=0.45\textwidth]{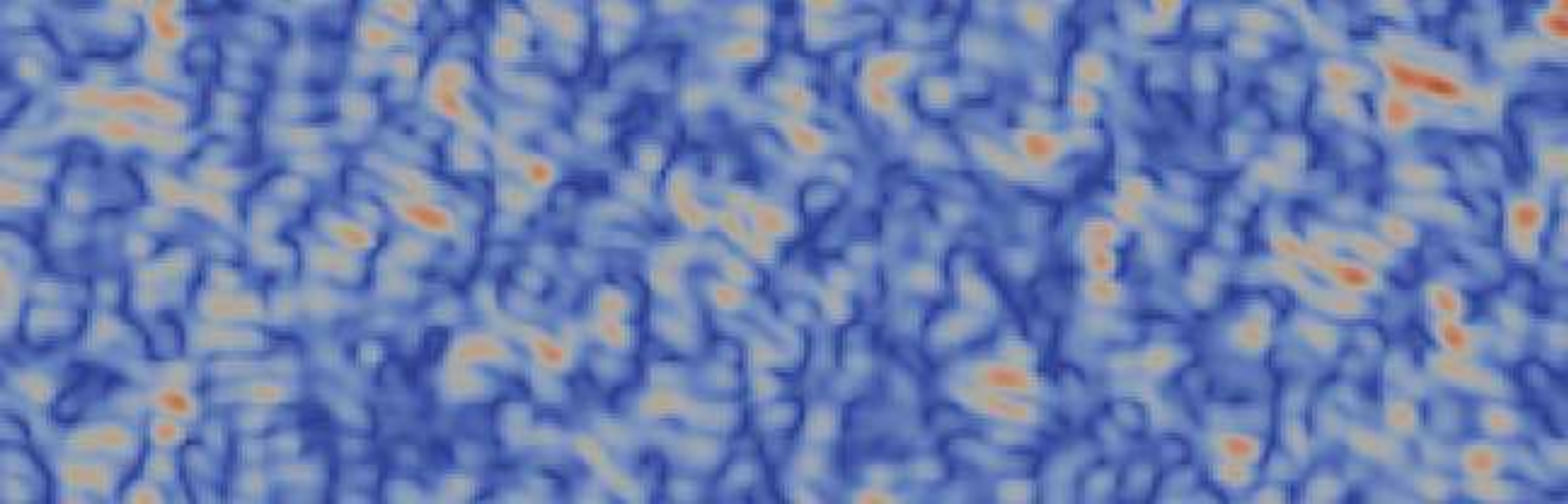}
 \includegraphics[angle=-90,width=0.45\textwidth]{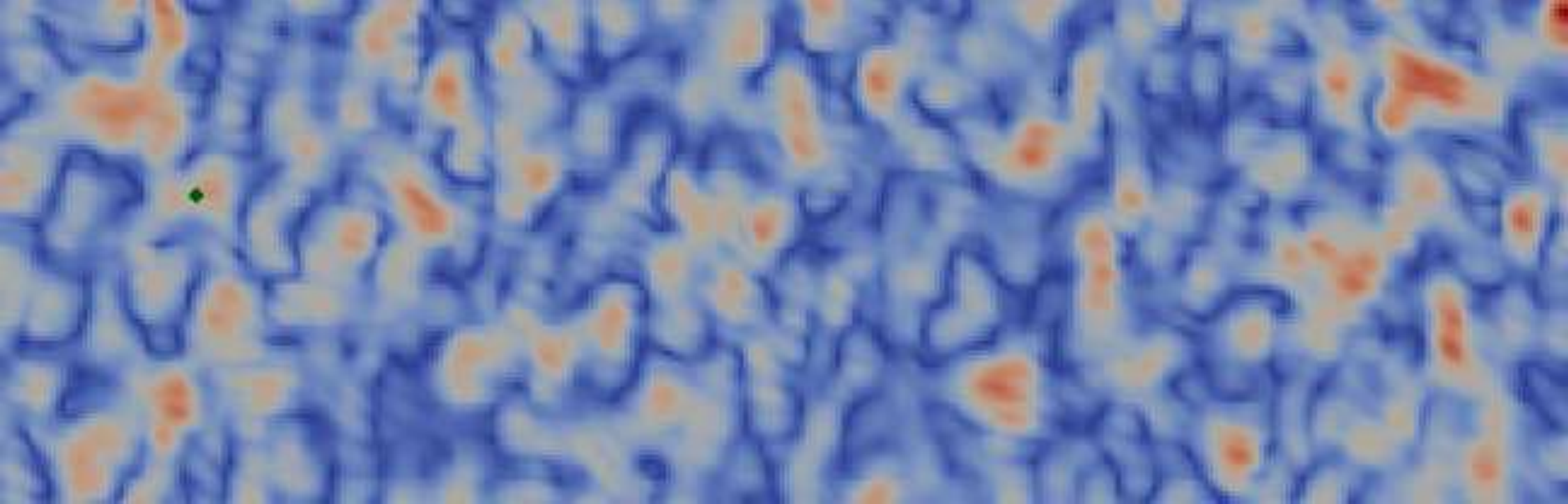}
 \includegraphics[angle=-90,width=0.45\textwidth]{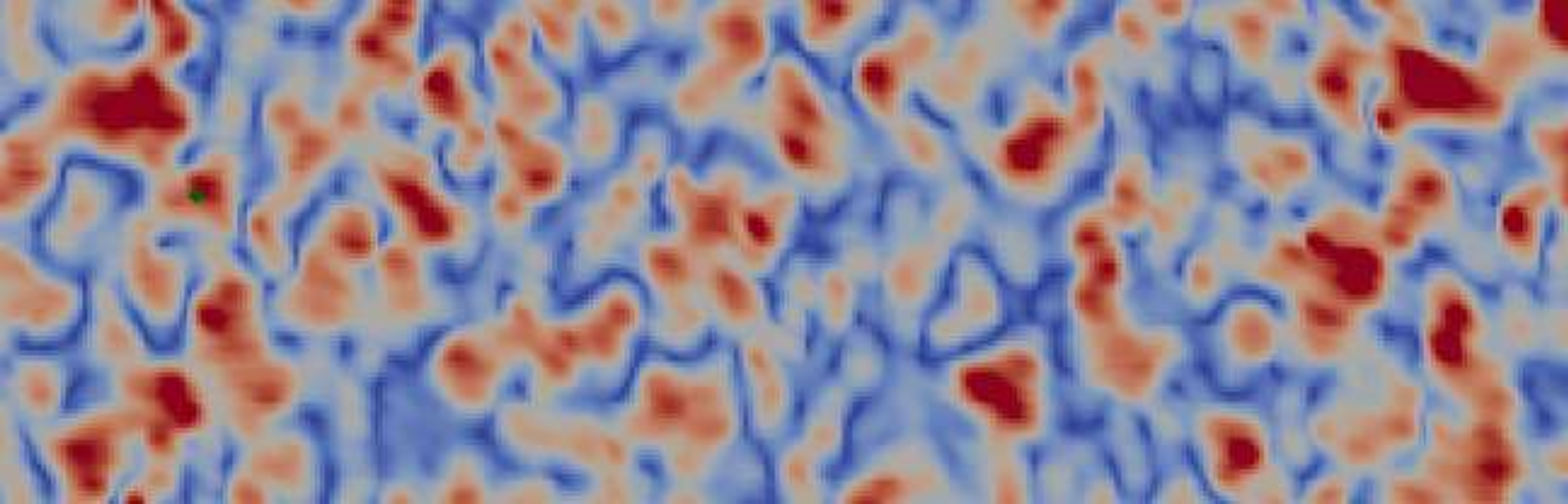}
 \includegraphics[angle=-90,width=0.45\textwidth]{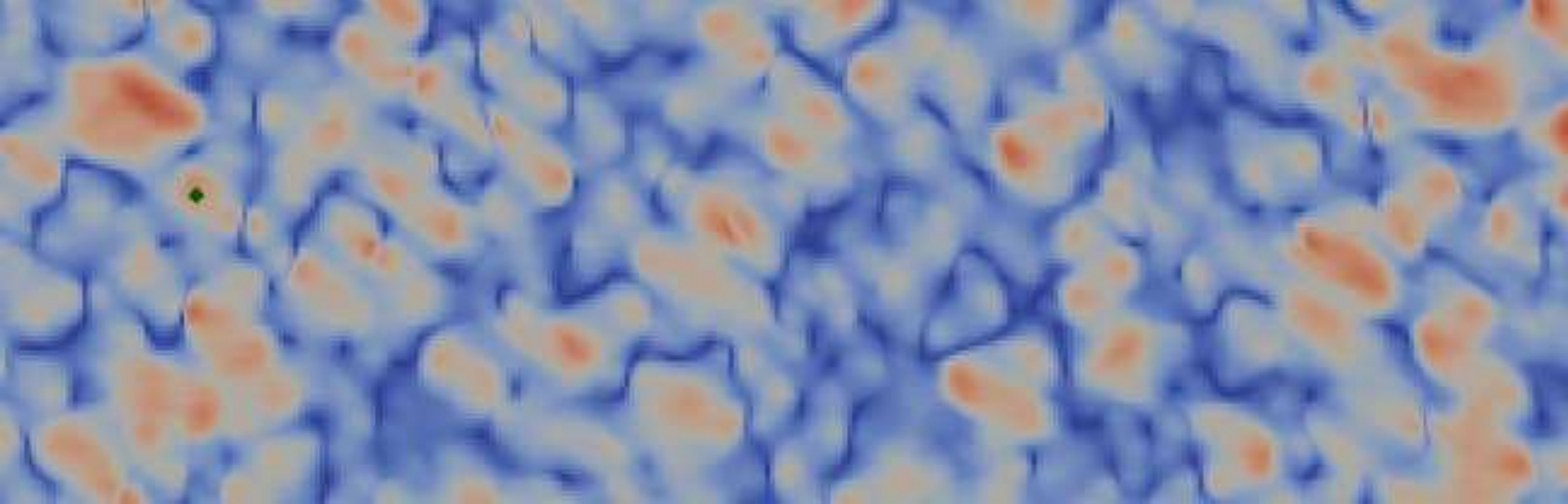}
 \includegraphics[angle=-90,width=0.45\textwidth]{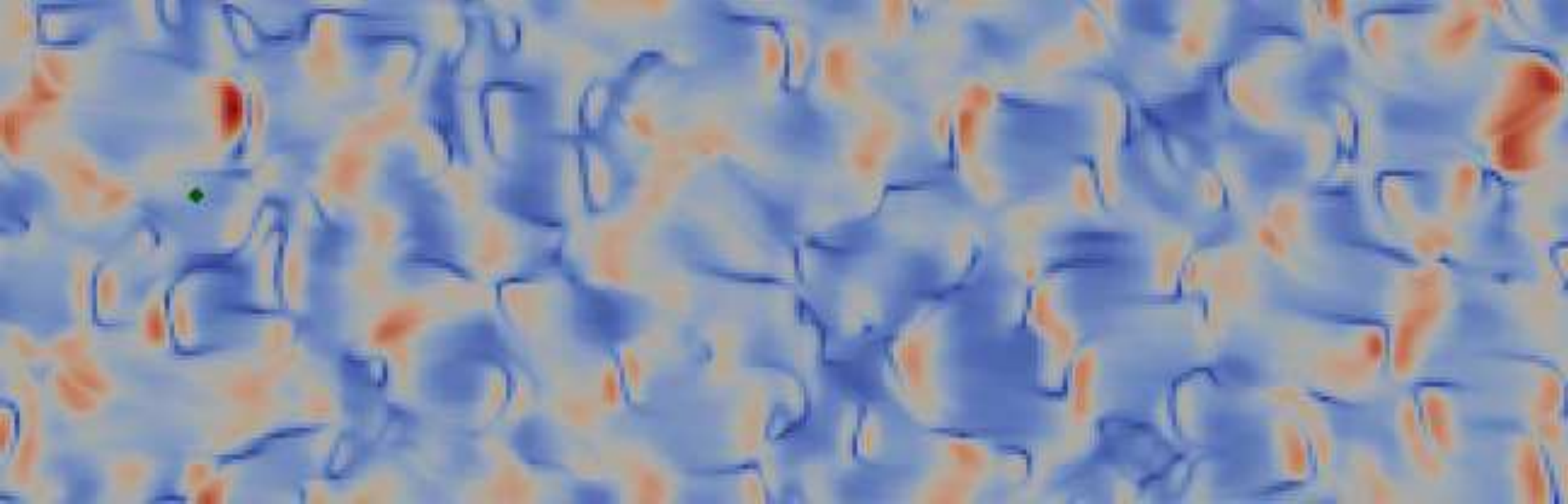}
 \includegraphics[angle=-90,width=0.45\textwidth]{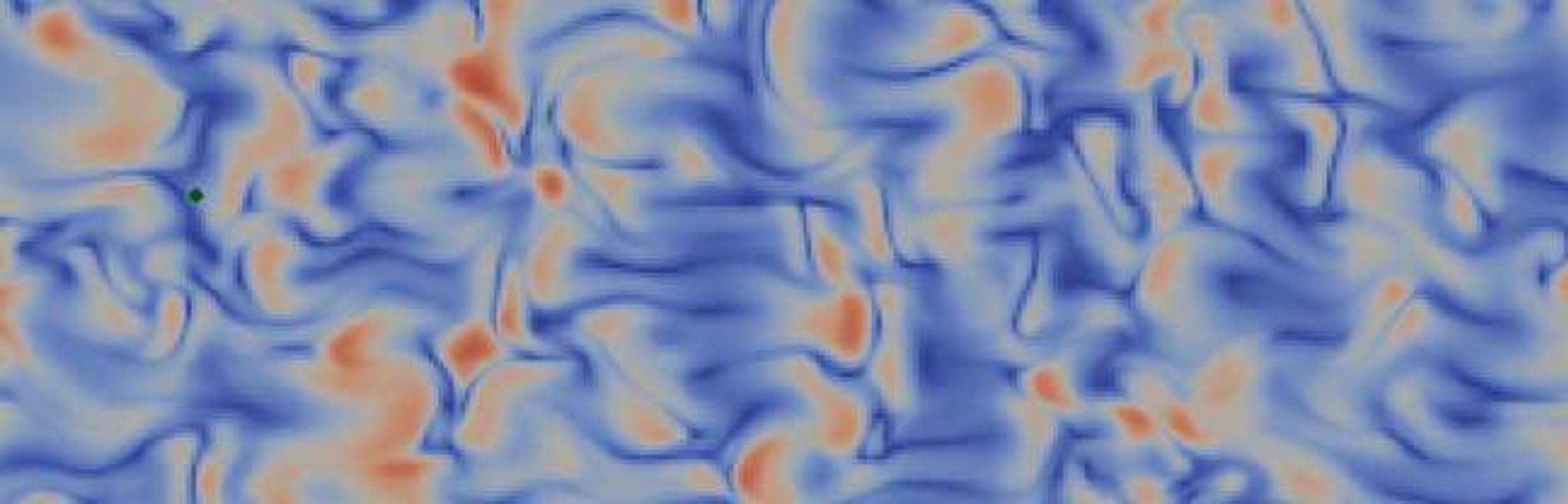}
 \includegraphics[angle=-90,width=0.45\textwidth]{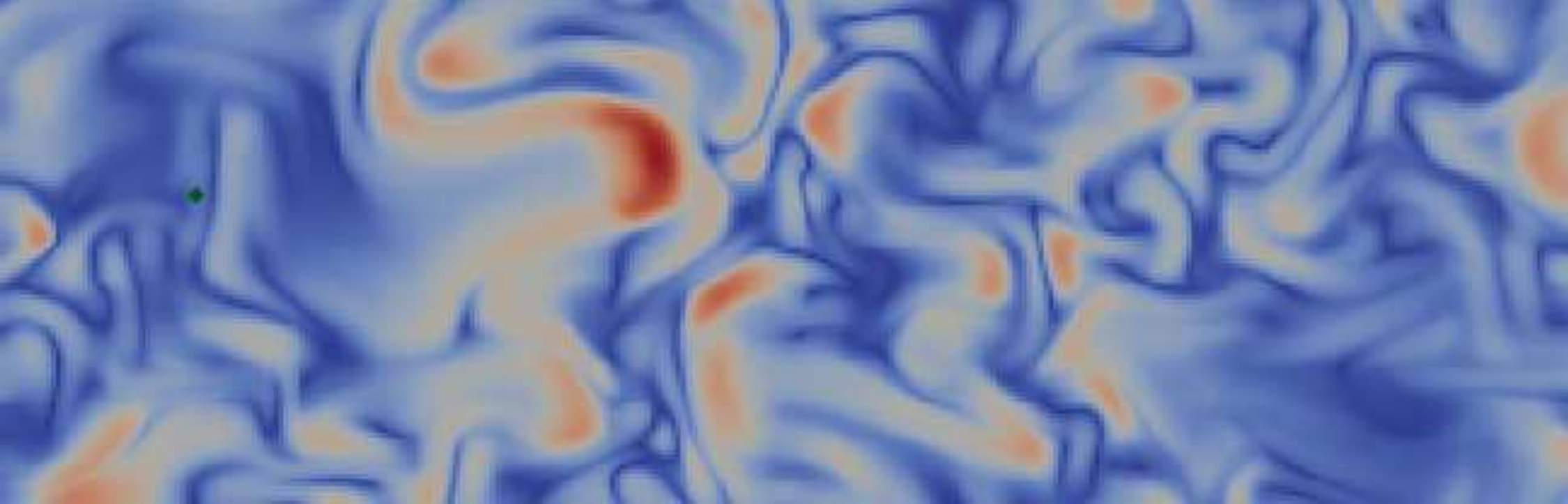}
  \caption[ ] {Evolution of $B_\perp/B_0$ (maximum displayed value between brackets) as a function of time, top to bottom: 0 (0.16), 16 (0.13), 32 (0.26), 48 (0.5), 63 (1.12), 79 (2.4), 103 (6.89), 127 (14.1), 158 (24.8) yrs ($B_{\perp,max}/B_0$). 
    \label{fig:fixedcurrent}}
  \end{figure}

In Fig.~\ref{fig:fixedcurrent}, we show the result of a simulation in which we use the following parameters: $\rho=2.34\times10^{-24}$~g~cm$^{-3}$, $B_0=5\;\mu$G, $j_0=5 \times 10^{-11}$~StA~cm$^{-2}$, and $t_{max}=5\times10^9$~s. The current density and $t_{max}$ are chosen such that $\gamma_{max}t_{max} \approx 10$ and significant amplification can be expected. The corresponding fastest growing wavelength is $\lambda_{min} \approx 3\times 10^{15}$~cm. In Fig.~\ref{fig:fixedcurrent} we use a resolution of $\Delta x\approx 2.4 \times 10^{14}$~cm and plot the value of $B_\perp/B_0$ as a function of time, with $B_\perp=\sqrt{(B_2^2+B_3^2)/2}$ perpendicular to the zeroth order magnetic field $B_0$. The cosmic ray current points from top to bottom, which is also parallel to $B_0$, and the width of the plots corresponds to $\approx 8.5 \times 10^{16}$~cm. The normalisation of the colour scale of each panel is different, the values are shown in the caption. 

Early on, best seen in panel 3 at $t=32$~yrs, a pattern is developing that has the shortest wavelength along the direction of the current and background magnetic field, as expected from the analytical theory \citep[e.g.][]{2005Bell}. The wavelength corresponds to the predicted value of $\sim 3 \times 10^{15}$~cm. As the fluctuations grow stronger, they grow in size and start merging. 
When $B_\perp/B_0$ starts to exceed unity, the clear initial distinction between parallel and perpendicular length scales starts to disappear, until in the 7th panel at $t\approx 100$~yrs little shocks are forming as the result of the perpendicular motion as the fluctuations grow in strength and scale. Growth slows down as the loops start to run into each other, and at very late times the original sense of direction of the instability is lost, as the loops start to swerve toward any direction of least resistance.

\begin{figure}
  \centering
 \includegraphics[width=0.45\textwidth]{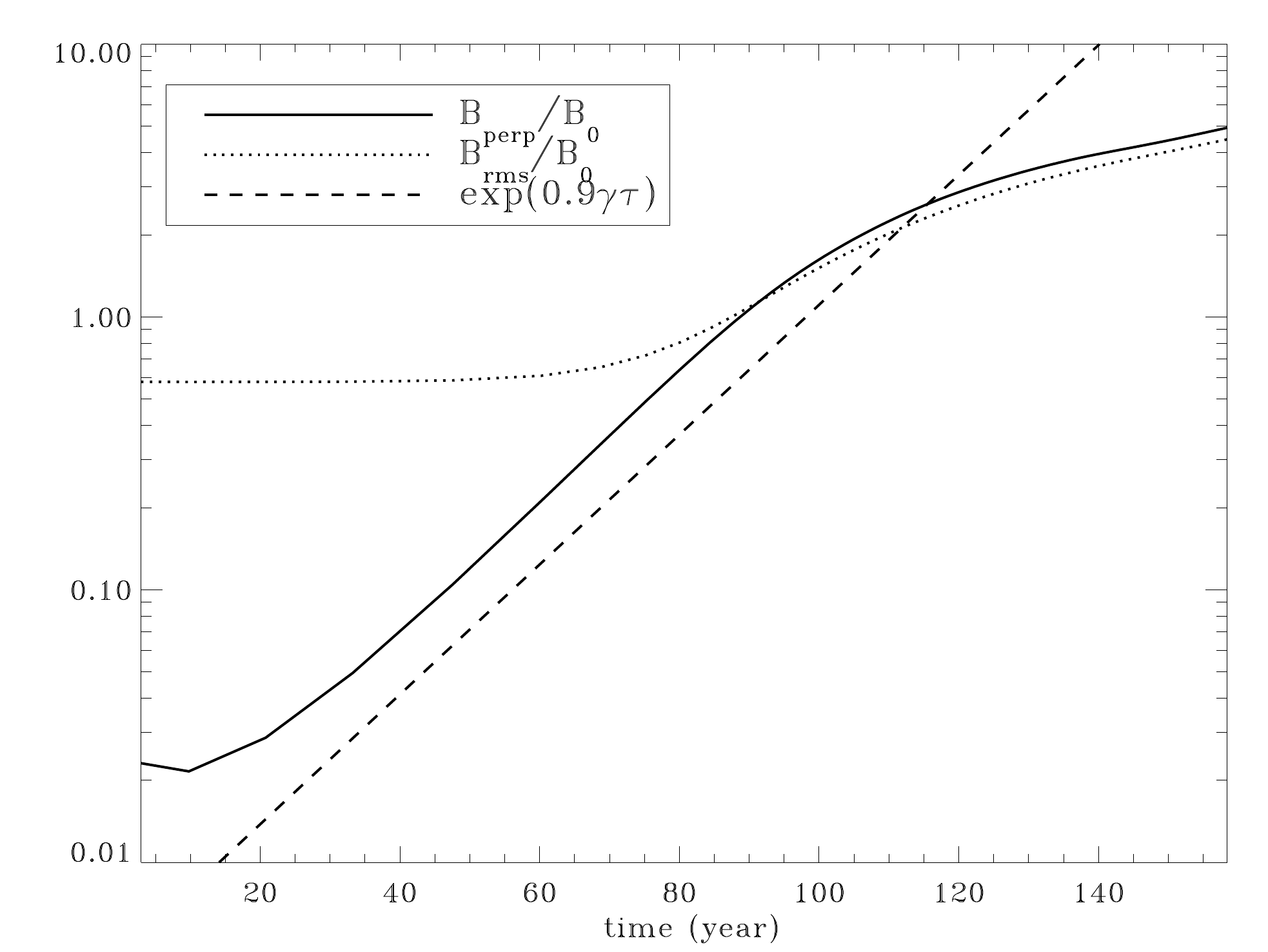}
 \includegraphics[width=0.45\textwidth]{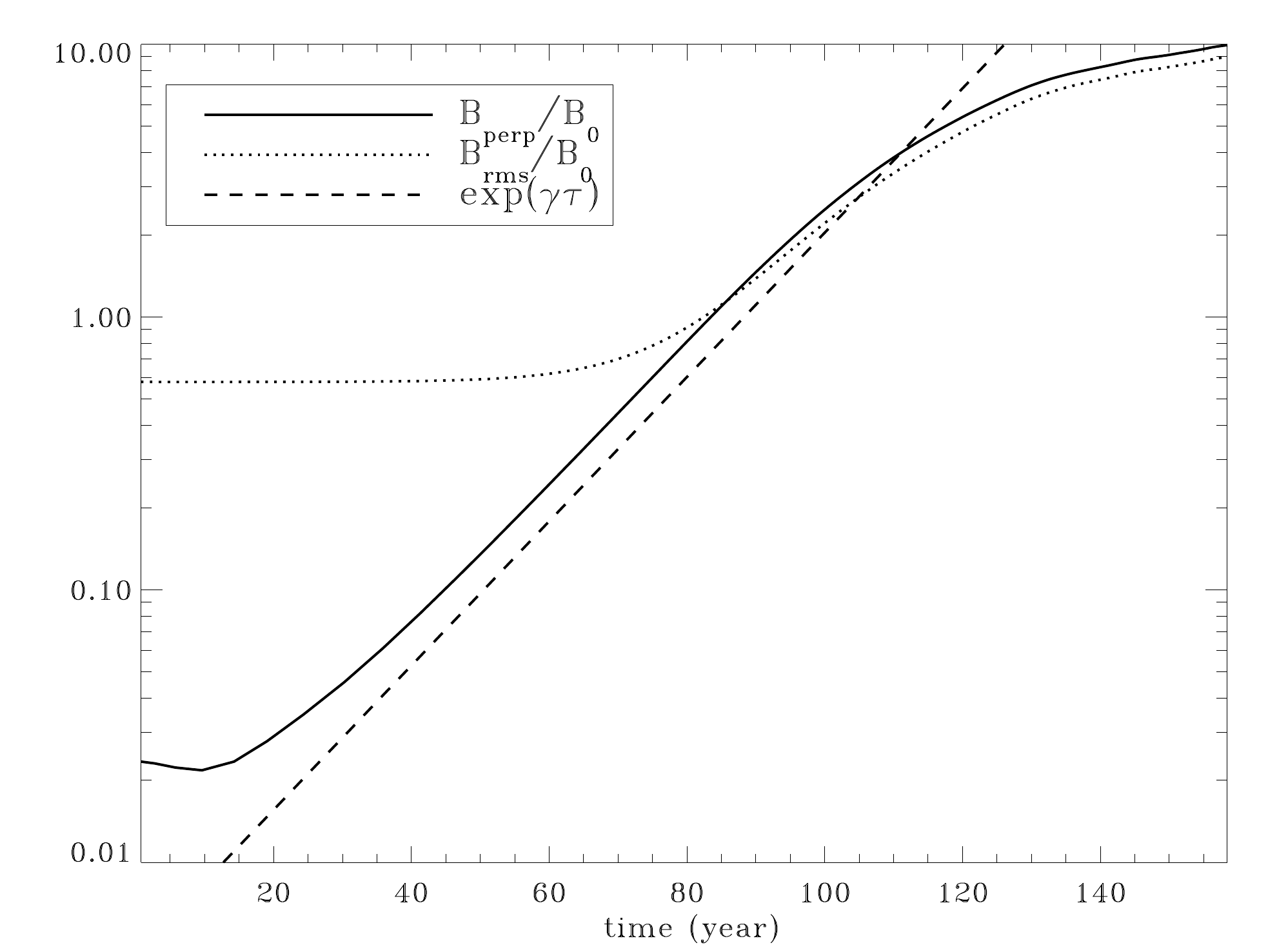}
 \caption[ ] {$\langle B_\perp \rangle$ (magnitude of $B_\perp$, averaged over the 2nd dimension) and  $B_{rms}$ as a function of time compared to the analytical growth rate. The lower plot has four times the resolution of the upper plot, and the growth rate in the highest resolution run is very close to the maximum value. 
    \label{fig:growthrate}}
  \end{figure} 
  
The growth rate of the magnetic field as a function of time is plotted in the top panel in Fig.~\ref{fig:growthrate} and is compared to $\langle B_\perp \rangle/B_0$ and $B_{rms}/B_0$ (with $B_{rms}=\sqrt{(B_1^2+B_2^2+B_3^2)/3}$). The growth rate is close to the analytical value, and seems to follow a slope of $\sim \exp(0.9 \gamma t)$. Doubling and quadrupling the resolution, such that $\Delta x\approx 6.2\times 10^{13}$~cm, brings the growth rate successively closer to the maximum value as derived from analytical theory (Eq.~\ref{eq:gammamax}). As can be seen in the lower panel of Fig.~\ref{fig:growthrate}, the highest resolution run generates magnetic field amplification very close to the maximum value. The higher resolution also allows for more compression, both in density and magnetic field, resulting in a higher value at which the magnetic field saturates. Initially, it takes a while for the magnetic field to start growing, which is due to some initial relaxation of the magnetic field, and the fastest growing mode needing some time to emerge from the seed fluctuations. The value at which the magnetic field starts to saturate however is not necessarily quantitatively reliable in these 2D simulations. 

We have run a 3D simulation to compare the quantitative and qualitative growth -- the results look very similar and therefore are not shown separately because they were done at half the resolution. The main difference between 2D and 3D runs is in the level at which the growth rate starts to reduce. In the nonlinear stage the growth in 3D is quenched more quickly, which is due to loops of amplified magnetic field running into each other, suppressing quick growth beyond that point, whereas in 2D one of the dimension is infinite and therefore does not slow down growth in that direction.

\begin{figure}
  \centering
 \includegraphics[angle=-90,width=0.225\textwidth]{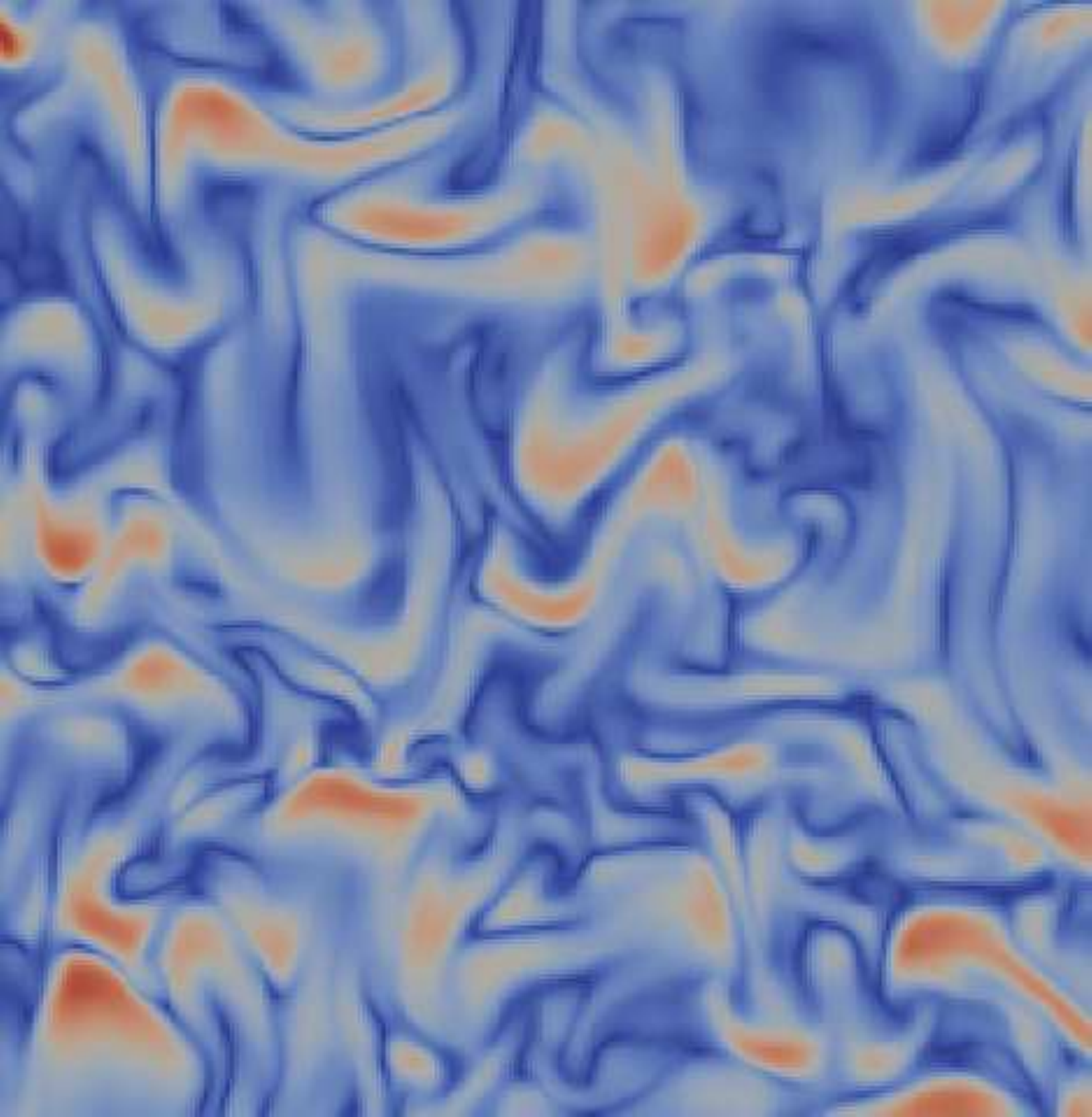}
 \includegraphics[angle=-90,width=0.225\textwidth]{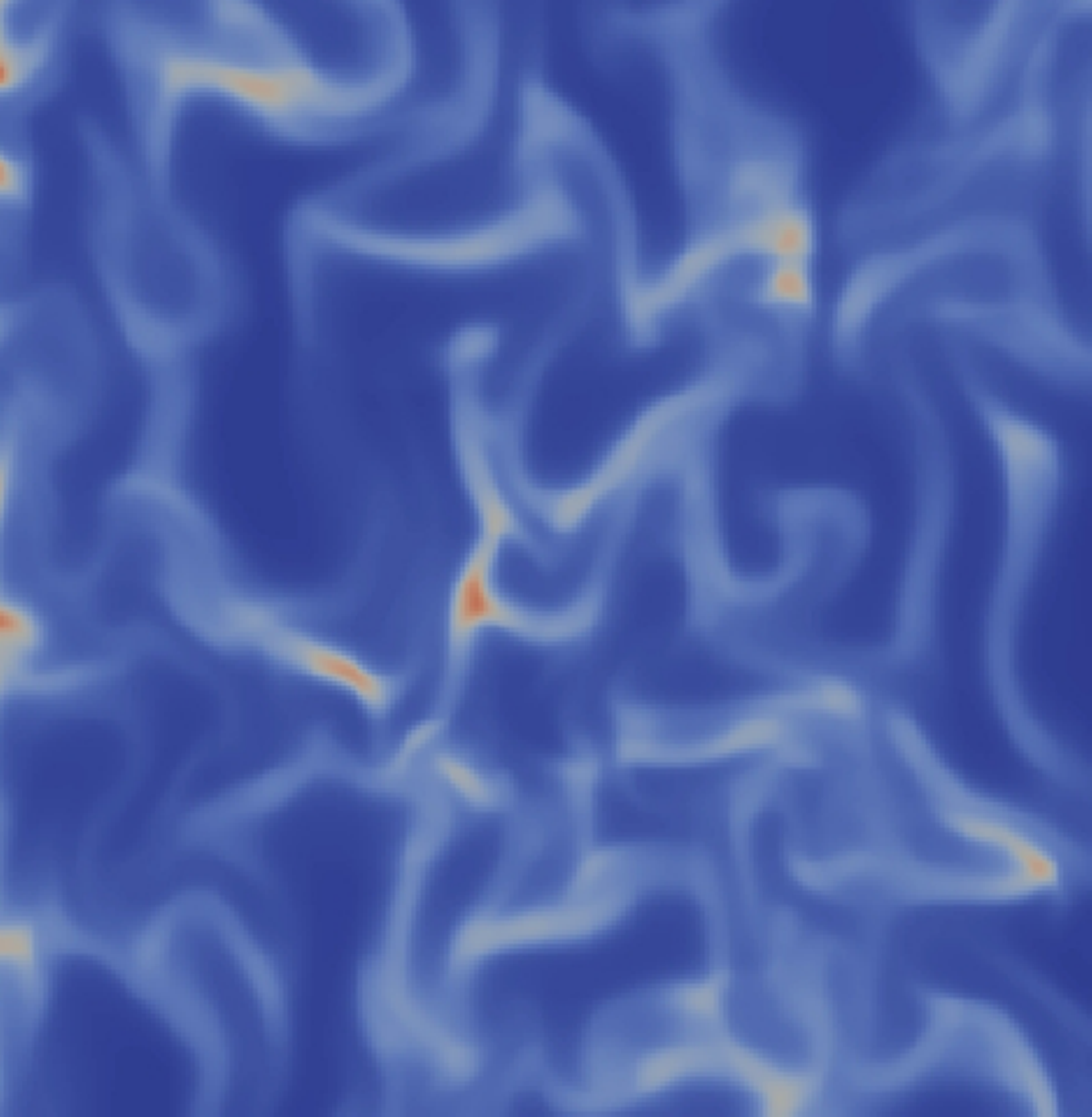}
 \includegraphics[angle=-90,width=0.225\textwidth]{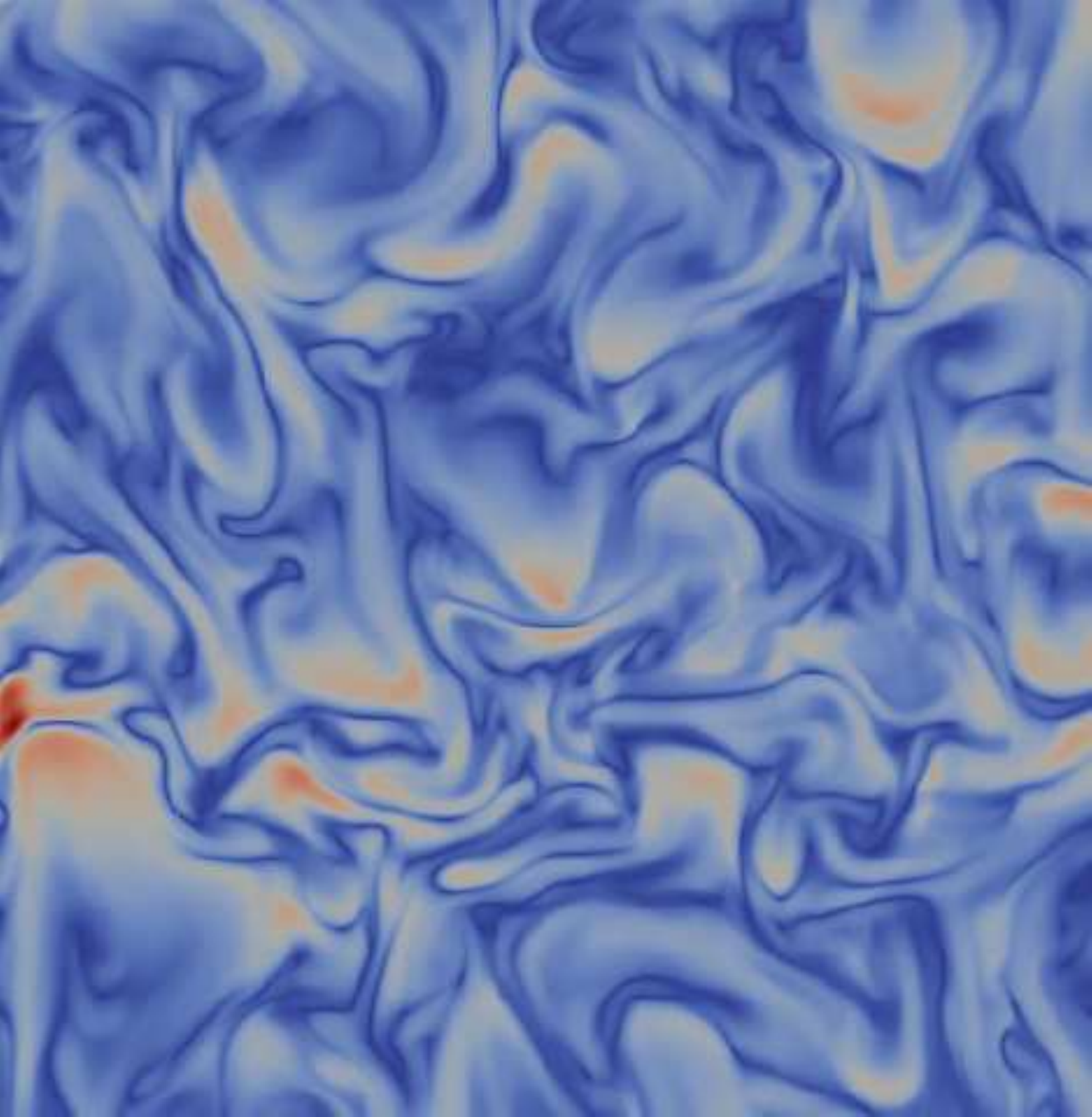}
 \includegraphics[angle=-90,width=0.225\textwidth]{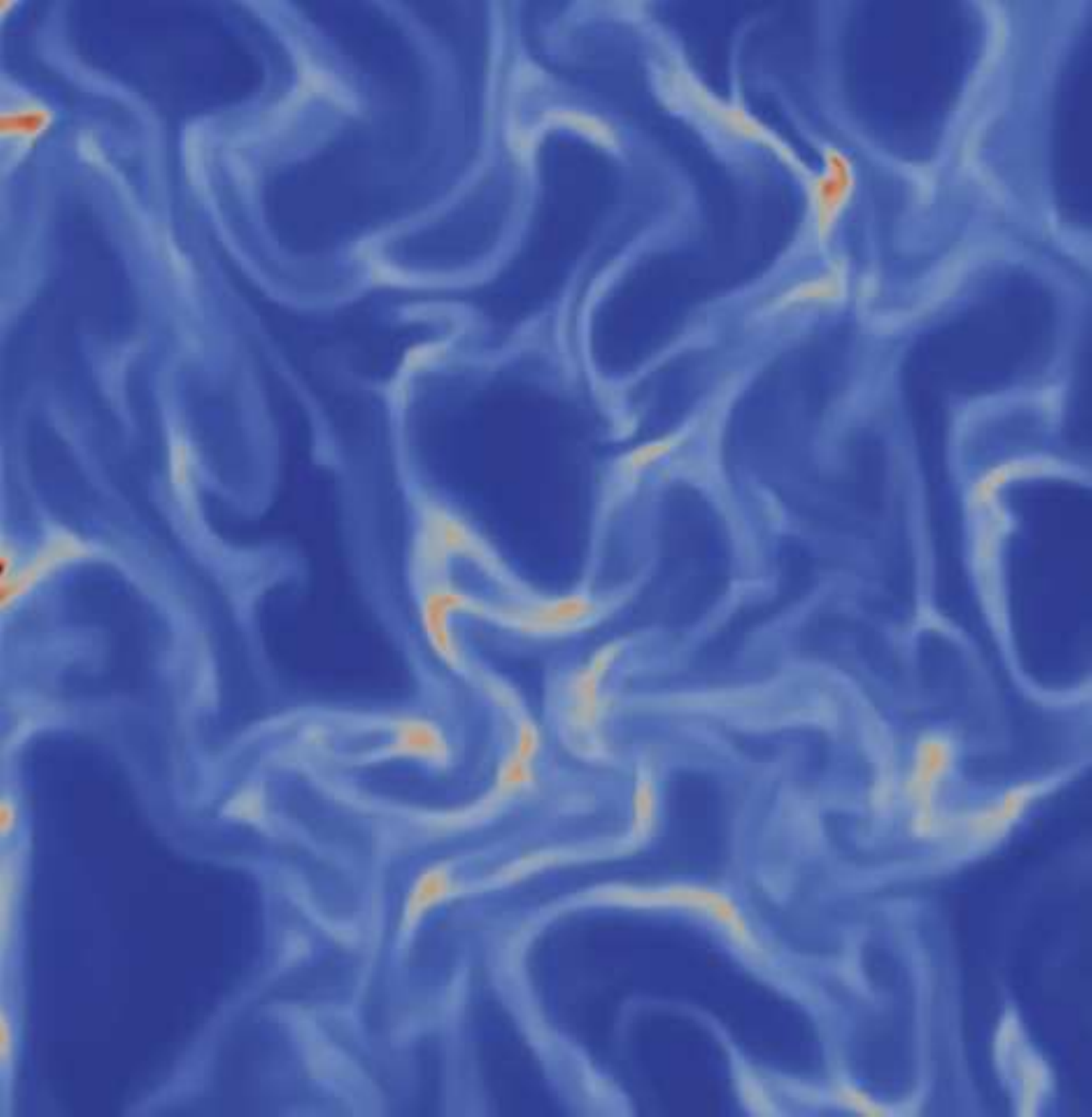}
 \includegraphics[angle=-90,width=0.225\textwidth]{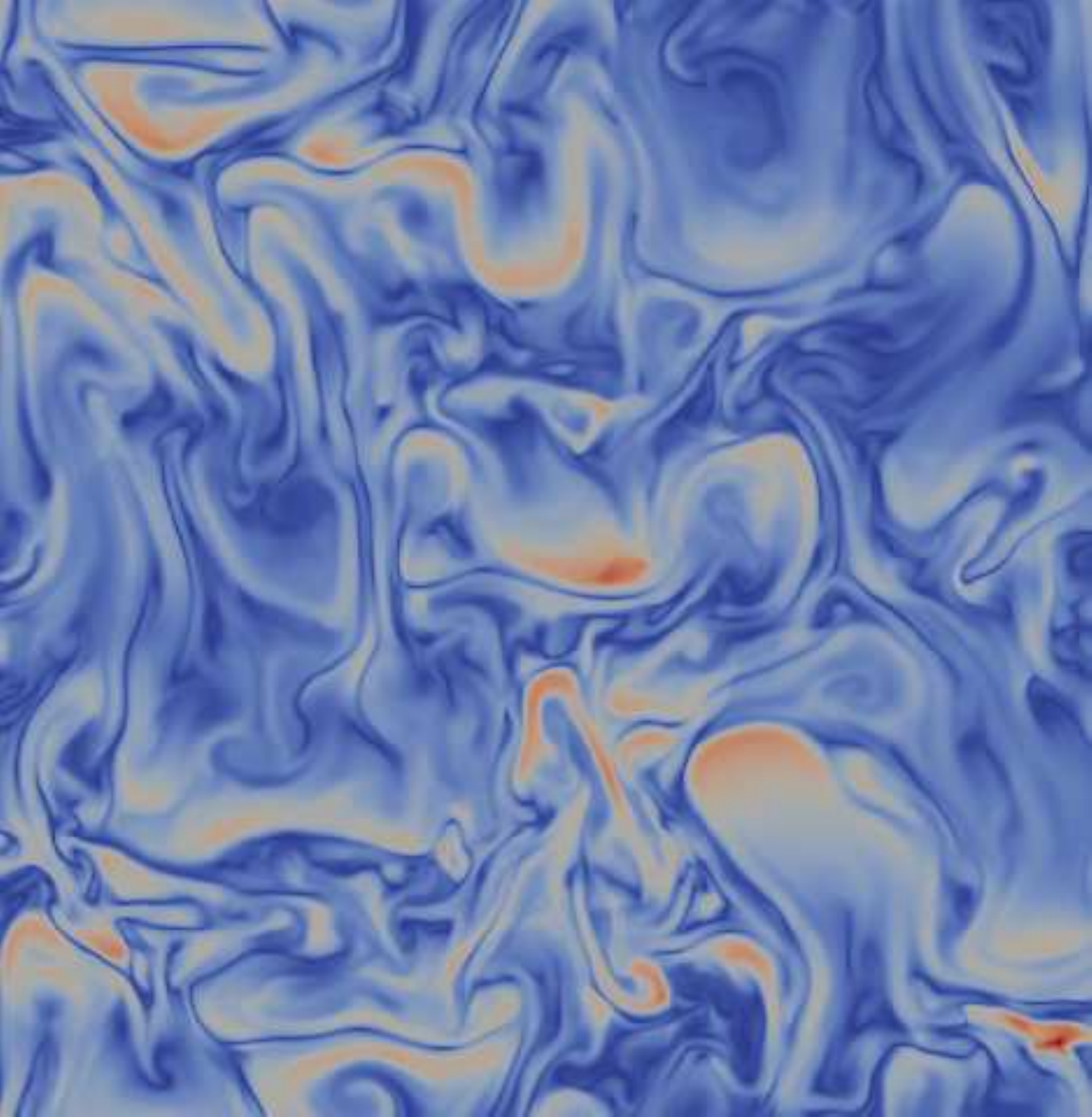}
 \includegraphics[angle=-90,width=0.225\textwidth]{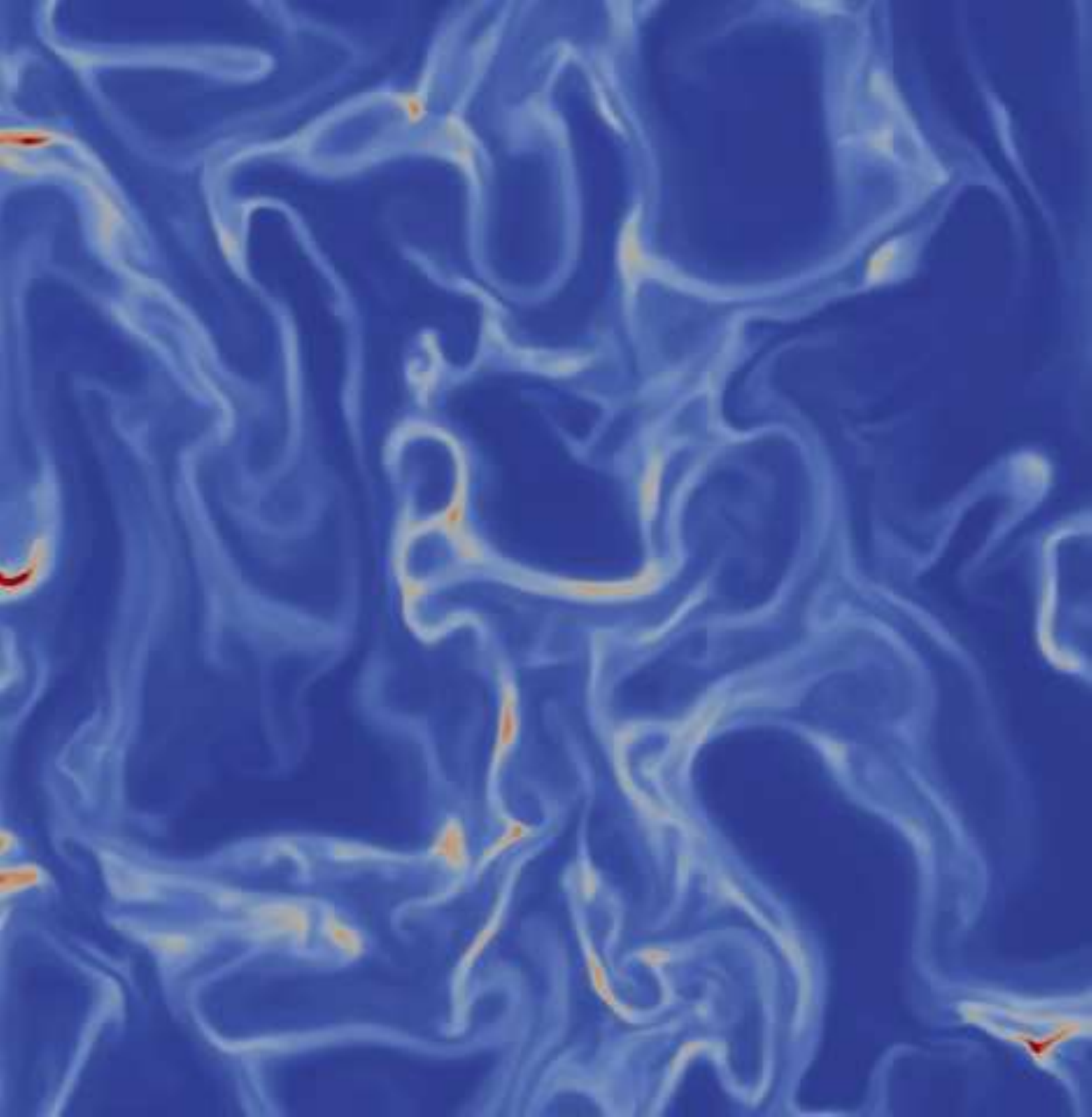}
  \caption[ ] {$B_\perp/B_0$ (left) and $\rho$ (right) at $t=158$~yrs for different resolution simulations. Although the overall features are the same, the growth rate of the instability is faster in the high resolution simulation and very close to the analytical value, whereas it is approximately 90\% of the analytical value in the lower resolution simulation. 
    \label{fig:resolution}}
  \end{figure}

In Fig.~\ref{fig:resolution} we show both $B_\perp/B_0$ and the density for the three different resolutions. Structurally, the low-resolution simulation does bring out all the features that are found in the higher-resolution simulation, but, as also could be seen from Fig.~\ref{fig:growthrate}, the growth rate is a bit lower than the maximum value. Similarly, the density contrast is higher when the resolution is higher.  
The resolution study shows that the lower-resolution simulation can still be used to get information on the magnetic field amplification and density structure, but for accurate quantitative statements a high resolution is necessary. For quantitatively reliable data in the non-linear phase 3D simulations are necessary. However, the trend, and growth rate up to $\langle B_\perp \rangle/B_0 \sim 2$ may well be studied in 2D with adequate accuracy. 
 
\section{Maximum energy}
\label{sec:Emax}

In the previous section we showed that the non-resonant instability acts as expected. After the amplified field becomes comparable to the background field, the growth slows down, but continues well into the non-linear regime. 
The growth of magnetic field fluctuations will aid in confinement of cosmic rays and therefore decrease the flux of escaping particles, i.e. decrease the escape current. In Sect.~\ref{sec:currentsource} we already derived how this requirement results in an estimate for the maximum cosmic ray energy that depends on mostly the shock parameters and our choice for $\gamma \tau$. Given a fixed fraction of the shock kinetic energy is transferred to the cosmic ray population, we subsequently derived maximum cosmic ray energy as a function of time as given by Eq.~\ref{eq:emax}.  
This energy is the maximum to which cosmic rays can be accelerated while leaving enough escaping particles to generate the cosmic ray current that drives the instability to amplify the magnetic field to a value such that $B_\perp \gg B_0$. 

In our derivation of Eq.~\ref{eq:emax}, we assumed the shock velocity was constant. In this section we expand this calculation by allowing for a variable shock velocity, while still taking into account the spherical dilution of cosmic rays upstream that is proportional $(R_s/R)^2$, and properly integrating the flux over time using a small numerical iteration routine that we describe below.

We calculate the evolution of the shock with 1D numerical simulations, where we initialise the ejecta with a flat density profile in the core and a powerlaw envelope \citep{1989ChevalierLiang}, with $\rho_{env} \propto r^{-9}$ for the core-collapse / wind scenario, and $\rho_{env} \propto r^{-7}$ for the ISM scenario. The velocity is initialised to be proportional with the radius, with a maximum value of 20,000~km/s. The ejecta mass and energy are respectively $M_{ej}=2.5$~M$_\odot$, $E_{ej}=2 \times 10^{51}$~erg (CSM) and $M_{ej}=1.4$~M$_\odot$, $E_{ej}=10^{51}$~erg (ISM). For the CSM case, we use a mass loss rate of $10^{-5}$~M$_\odot/yr$ and wind velocities of $4.7$, $15$, and $1000$~km/s, typical for Type II and Type Ib/c SNe that explode in either a (dense) red supergiant (RSG) wind, or a tenuous Wolf-Rayet (WR) wind. We evaluate the ISM case for two different interstellar densities $n_{ism}=0.05$ and $n_{ism}=0.85$. 

With our variable shock velocity from 1D hydrosimulations, we use a small numerical iteration routine to get a better estimate for $E_{max}$ by using our shock parameters in Eq.~\ref{eq:gt_int}, and using Eq.~\ref{eq:j_pz} for the current density. As we discussed previously, which value we have to take for $\gamma \tau$ is up for discussion and should be somewhere between $\ln(B_0/B_{\perp,0}) < \gamma \tau \leq \ln(B_{sat}/B_{\perp,0})$. We compare our calculations with observations to obtain a best first estimate for $\gamma \tau$. 

As summarised in Table~\ref{table:models}, the observed values for Cassiopeia A (Cas A) and Tycho give us a magnetic field value of the order of $B \approx 200\;\mu$G, representing the downstream (post shock) field \citep[e.g.~][]{2012Helderetal}. Assuming that the compression due to the shock is about $\sqrt{11}$, this gives for the upstream value $B \approx 60 \;\mu$G. Continuing with our model assumption, that initially $B_0 \approx 5 \;\mu$G, and the seed fluctuations are on the order of 5\%, we get an estimate of $\gamma\tau=\ln(60\;\mu$G$/(0.05 \times 5\;\mu$G$))\approx5.5$. We will use a value of $\gamma \tau=5$ for convenience (if we had taken fluctuations on the 10\% level this value would have come out naturally). If we look at the growth rate in Fig.~\ref{fig:growthrate}, we find that at $\gamma \tau=5$, ($t=83$~yr), $\langle B_\perp\rangle/B_0$ has reached a value of about unity. If growth would start more rapidly initially, growth could potentially be larger at this time. So, as a first approach, $\gamma \tau=5$ from multiple aspects seems to be a reasonable starting point. 

\begin{figure*}
  \centering
\includegraphics[width=0.49\textwidth]{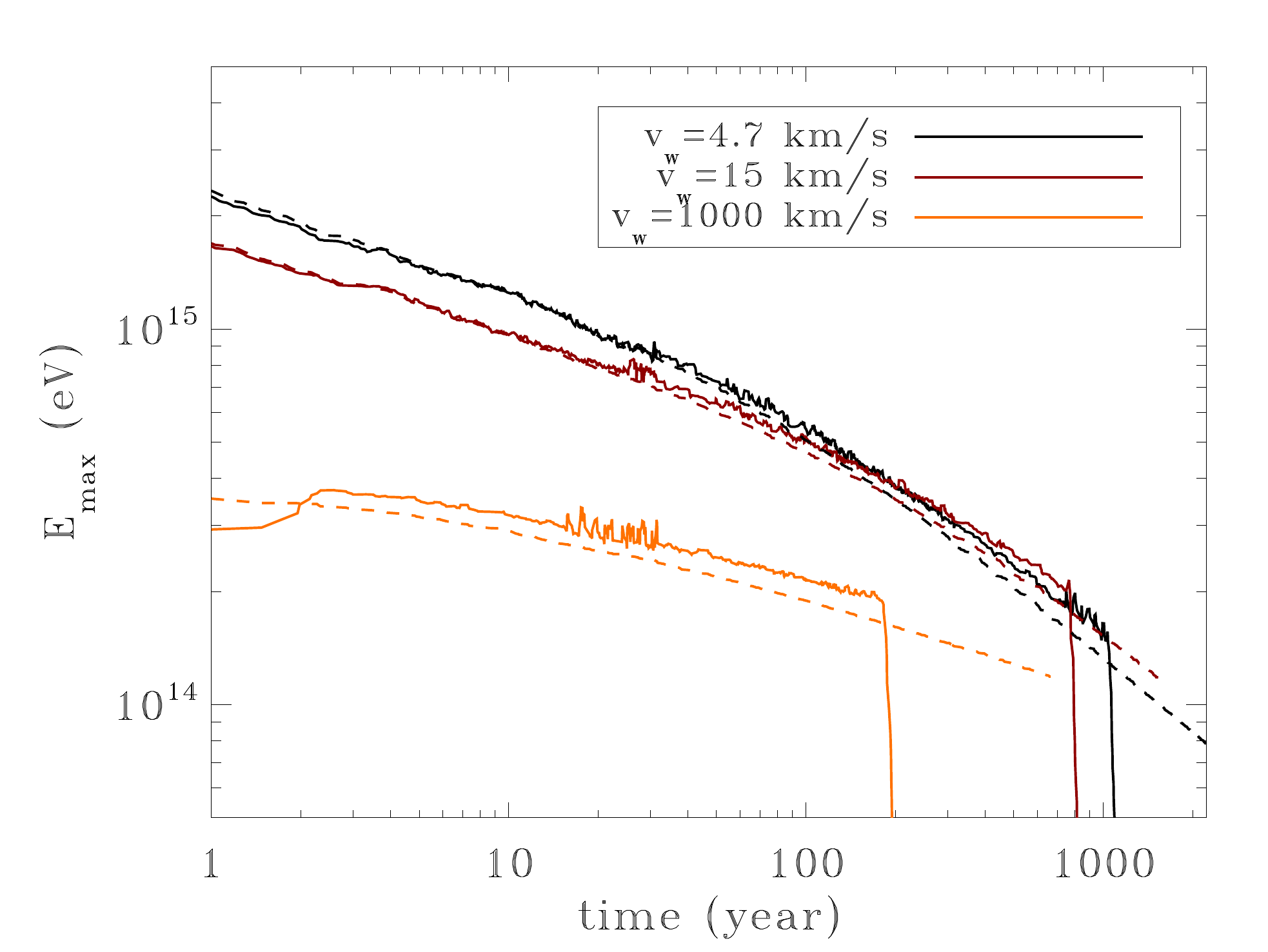}
\includegraphics[width=0.49\textwidth]{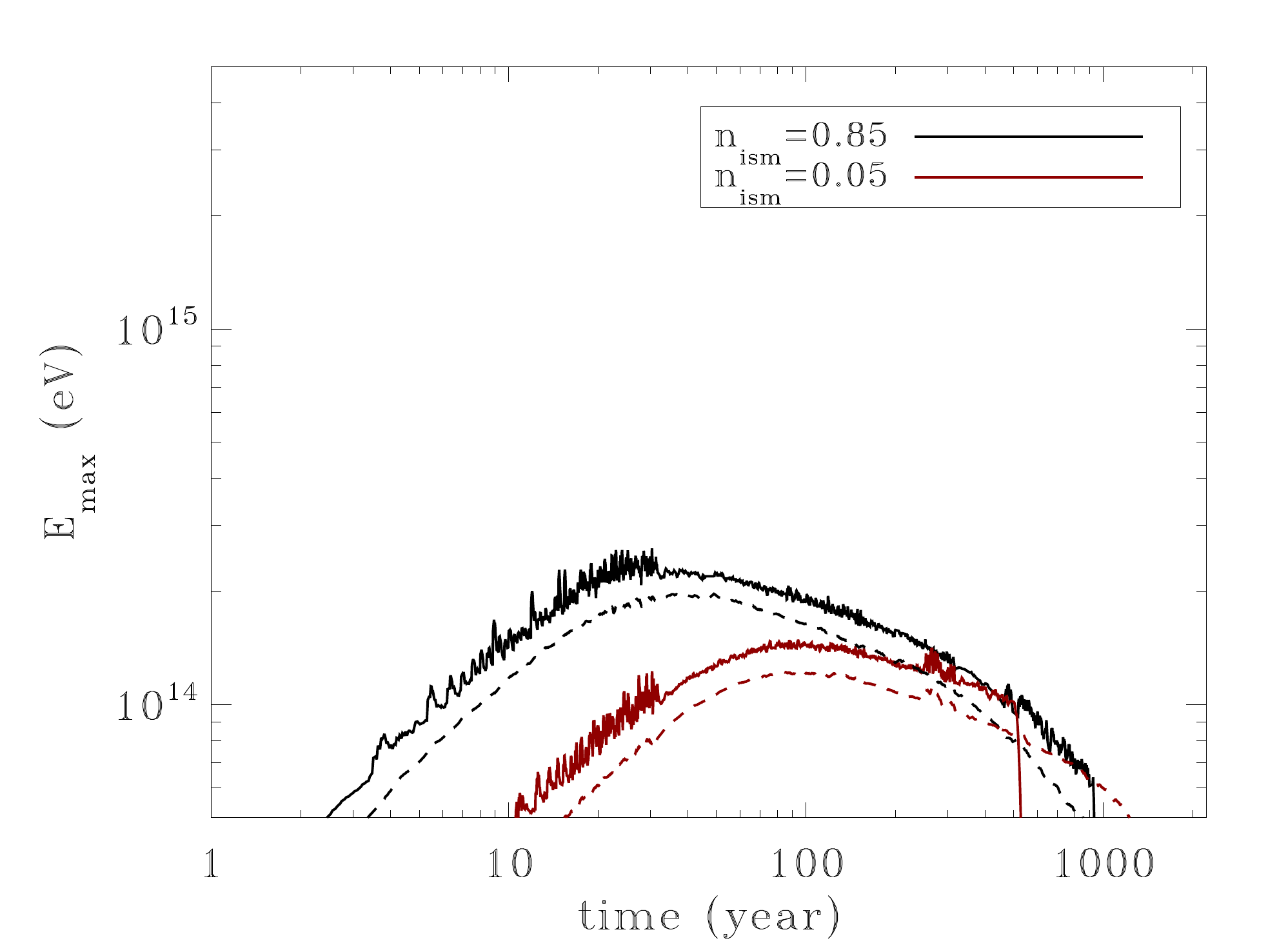}
 \caption[ ] {Maximum energy as a function of time for the evolution of a SNR in various environments. Left: The black, red, and yellow curves indicate $E_{max}$ for a CSM created by wind velocities of $4.7$, $15$, and $1000$~km~s$^{-1}$ respectively, and a mass loss rate of $\dot M=10^{-5}$~M$_\odot$ in all three cases.  Right: The black and red curve indicate $E_{max}$ for an ISM with a number density of $0.85$ and $0.05$~cm$^{-3}$ respectively. The dashed line shows the analytical solution given by Eq.~\ref{eq:emax}, using $\ln(E_{max}/m_p c^2)=14$, whereas the solid line shows the numerically integrated solution for the maximum energy that takes into time dependence of the shock velocity. The steep drop is where the NRH instability stops being effective and where other instabilities will be required to grow the magnetic field fluctuations.}
    \label{fig:Emaxall}
 \end{figure*}

The numerical iteration that we use to obtain $E_{max}$ as a function of time, now allowing for a variable shock velocity, is as follows. We set $\gamma \tau=5$ in Eq.~\ref{eq:emax}, to get a first estimate for $E_{max}$, and, because we are looking for the maximum possible $E_{max}$ that is still consistent with our requirement on $\gamma \tau$, we use 1.5 times this value as the starting point for our numerical iteration. We then solve the following: {\em 1):} From $E_{max}$ we derive a provisional current using Eq.~\ref{eq:j_pz}. {\em 2):} This current is used in Eq.~\ref{eq:gt_int} to calculate the value of $\gamma \tau$ as a function of radius, taking into account that the current decreases with $R^2$. {\em 3):} We lower $E_{max}$, and go back to step 1 until the current is sufficient such that the integrated value of $\gamma \tau$ from Eq.~\ref{eq:gt_int} at the shock radius is larger than $5$. Also, we check that the gyroradius in the unamplified magnetic field is larger than the wavelength of the smallest growing mode of the NRH instability, for otherwise this approach is not valid and we set $E_{max}=0$. Finally, we compare $E_{max}$ that comes out of this iteration with the analytical result from Eq.~\ref{eq:emax} and plot them as a function of time for the various SNR evolution scenarios in Fig.~\ref{fig:Emaxall}. 

\begin{figure*}
  \centering
\includegraphics[width=0.49\textwidth]{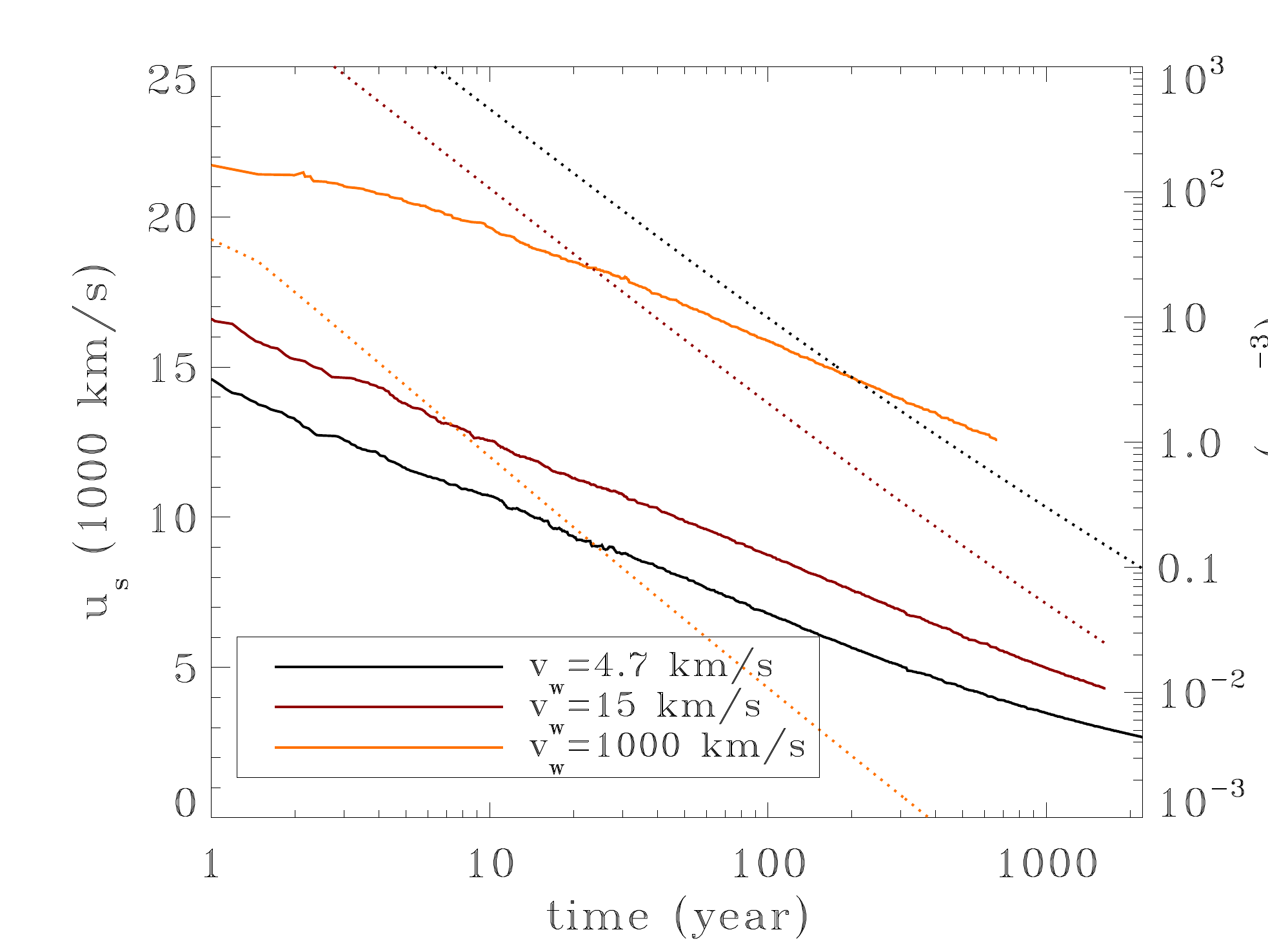}
\includegraphics[width=0.49\textwidth]{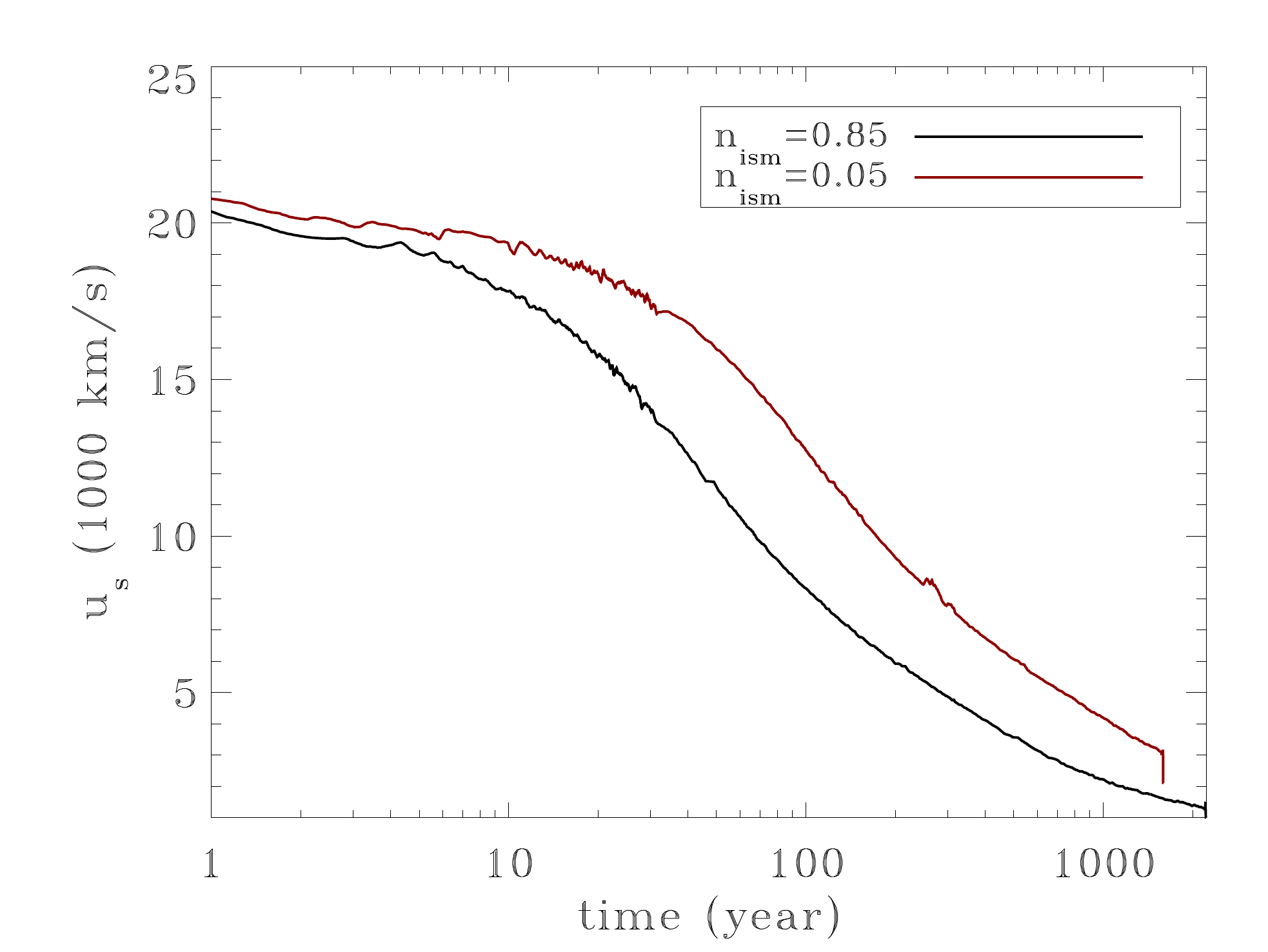}
 \caption[ ] {Shock velocity as a function of time for the evolution of a SNR in various environments. Left: CSM for various values of the wind velocity, as in Fig.~\ref{fig:Emaxall}. The solid lines indicate the shock velocity and the dotted lines give the number density just upstream of the shock as a function of time. Right: evolution of the shock velocity for the two different values for the ISM number density. }
    \label{fig:usall}
 \end{figure*}

The calculated numerical values for $E_{max}$ as a function of time are plotted in Fig.~\ref{fig:Emaxall} for SN ejecta evolving in a circumstellar medium (CSM: wind profile density) and in a homogeneous interstellar medium (ISM).
The RSG parameters used are representative for a Cassiopeia A type of explosion, and for the Type Ia environment we have used parameters representative for Tycho ($n_{ism}=0.85$ \citep{2006Badenesetal}, but for a discussion see \citet{2013Chiotellisetal}), and for SN 1006 ($n_{ism}=0.05$). In Fig.~\ref{fig:Emaxall} the solid lines show the numerically integrated results, and the dashed curves show the analytical results from Eq.~\ref{eq:emax} with $\gamma \tau = 5$ and $\ln(E_{max}/m_p c^2)=14$. We find that at early times ($t \leq 10$~yr) there is a deviation between the analytical solution and the numerical, owing to the instability needing time to grow. Also, for low energies, the analytical solution is lower than the numerical solution because $\ln(E_{max}/m_p c^2)$ drops below 14. For reference, we have also plotted the shock velocity for the various models, as resulting from our 1D hydro simulations, in Fig.~\ref{fig:usall}.

Given the numerical and analytical value are very close, we can conclude that Eq.~\ref{eq:emax} provides a good representation of the maximum cosmic ray energy as a function of time. Our chosen values for $\chi=0.34$ and $\gamma \tau=5$ can be adjusted if the observations provide us with information to justify so.

The values for $E_{max}$ that result from our numerical analysis for the current ages of the SNRs are summarised in Table~\ref{table:models}, along with the shock radius and velocity as from our 1D hydrodynamical models, and the calculated and observed values for the magnetic field strength. Observations will measure the post-shock field rather than the upstream field, which is the one we calculate. For compression of a factor 4 of a completely turbulent field, the post-shock field is expected to be greater than the pre-shock field by a factor $\sim \sqrt{11}$. That $\sim \sqrt{11} B_{sat} > B_{obs}$ for Cas A and (albeit less so) for Tycho, indicates
that magnetic field amplification is limited by growth times rather than saturation in these cases. For our SN1006 model, the value for $E_{max}$ is derived directly from Eq.~\ref{eq:emax} as an upper limit, since with our used model parameters the NRH instability is quenched after about $500$~yrs. Similarly, the corresponding magnetic field is regarded as an upper limit.

Note that the value quoted for $E_{max}$ is for $\gamma \tau=5$, which if we assume exponential growth, results in a magnetic field value of $0.05 \times 5 \times e^{5}=37\;\mu$G. Even a small change in $E_{max}$ will lead to a large difference for the magnetic field value. An increase in $\gamma \tau$, for example from 5 to 6, would increase the upstream magnetic field from $\sim 37$ to $\sim 100\;\mu$G, while only decreasing $E_{max}$ by $\sim$17\% compared to the value quoted in Table~\ref{table:models}. Potentially, $\gamma \tau$ is variable and depends on the saturation criterion -- if the magnetic field were amplified to the value quoted as $B_{sat}$, the corresponding $E_{max}$ would be lower by a factor of $\ln(B_{sat}/B_{\perp,0})/5$, which is equivalent to replacing a $\gamma \tau$ of $5$ with a $\gamma \tau$ of $\ln(B_{sat}/B_{\perp,0})$.

\begin{table*}
\centering                        
        
\begin{tabular}{c  c  c   c   c  c  c}       
\hline\hline
SNR Type & age & $R_s$ & $u_s$ & $E_{max} \left(\gamma\tau=5\right)$  & $B_{sat}$ & $B_{obs}$
\\
\hline
RSG (Cas A) & 330 yr & 2.2 pc & 4900 km s$^{-1}$ & 283 TeV &  $243\;\mu$G & $210-230\;\mu$G \\ 
Tycho & 440 yr & 3.2 pc & 3900 km s$^{-1}$ & 108 TeV &  $128\;\mu$G & $200-230\;\mu$G \\
SN1006 & 1000 yr & 7.6 pc & 4100 km s$^{-1}$ & $<60$ TeV  & $<35\;\mu$G & $80-150\;\mu$G \\
\hline
\end{tabular}
\caption{\textrm{Overview of the supernova remnant models, evaluated at times that correspond to the age of their closest corresponding real SNR. $R_s$ and $u_s$ are from the 1D hydrodynamical models, and the values of $E_{max}$ are from the numerical iteration routine that solves Eq.~\ref{eq:gt_int} for variable shock velocity (solid lines in Fig.~\ref{fig:Emaxall}), which is close to what you would get directly from Eq.~\ref{eq:emax}. 
$B_{sat}$ is the calculated saturation value for the magnetic field immediately upstream of the shock
as derived from Eq.~\ref{eq:Bsat}. 
$B_{obs}$ is the post-shock field deduced from X-ray synchrotron observations \citep[e.g.][]{2012Helderetal}.
The post-shock field is expected to be greater than the pre-shock field by
a factor $\sim \sqrt{11}$.}  \label{table:models}}    
\end{table*} 

For both Tycho and Cas A downstream magnetic field values are estimated to be in the range $B>200-230\;\mu$G. For a compression of about $\sqrt{11}$ at the shock, barring additional amplification that could piggy-back of the NRH instability, this would imply $B \approx 60\;\mu$G for the upstream field, or $\gamma \tau \approx 5.5$, lowering $E_{max}$ to about 270 TeV for Cas A, and approximately 96 TeV for Tycho. From $\gamma$-ray observations that include detections of energies above 10 TeV by VERITAS for Tycho \citep[][]{2011Acciarietal} and 20 TeV for Cas A \citep{2001Aharonianetal,2010Acciarietal} by HESS and VERITAS, a characteristic proton energy of 7 times that value is not far off from the saturation value we find in our model. The high magnetic field will cause severe synchrotron losses at the high energy end of the electron spectrum. For more detailed models on how an ad-hoc cosmic ray spectrum translates to the photon spectrum, we refer to more detailed models in this respect regarding Tycho's SNR by \citet[e.g.][]{2013Berezhkoetal,2012MorlinoCaprioli}. There is also still uncertainty on the distance estimates of the observed galactic SNRs. For example in the case of Tycho there is a discrepancy between the expected environment based on dynamics and that based on ionisation age. Because the distance is not well-known, the radius and shock velocity are uncertain. The density of the environment has been estimated to be between values of $n_{ism}\approx 0.05-0.85$, or, to match both dynamics and ionisation age a more complicated CSM could be required \citep{2013Chiotellisetal}.

For the SN1006-like model, using our numerical integration procedure, we find that the NRH instability is no longer operating after about 500~yrs, when the gyroradius of the highest-energy cosmic rays is smaller than the wavelength of the fastest growing mode. 
Given that our numbers underestimate the radius and velocity somewhat, this may not necessarily be the case in SN1006 and a more targeted study is needed to determine if the NRH could still be operating in the system.
It is possible that once the NRH stops being effective, another instability, e.g. the resonant one \citep{1969KulsrudPearce,1974Wentzel}, becomes dominant and carries on the amplification and acceleration process, albeit probably at a slower rate. This seems to be a possibility for the current situation in SN1006, especially considering most of the observational estimates for the magnetic field are not very high. Magnetic field values found in the literature, often from fitting models to the observational data for SN 1006, cover a range of downstream values of $80\;\mu$G, $> 60-90\;\mu$G, $120\;\mu$G, $150\;\mu$G,  \citep{2012ArayaFrutos, 2006Parizotetal, 2010Aceroetal,2012Berezhkoetal}, and a model by \citet{2011Petruketal} quotes a preferred upstream value of $25\;\mu$G ($12\;\mu$G if no protons are accelerated). It seems that in SN1006 amplification does indeed proceed much slower than in e.g. Tycho and Cas A. 

However, HESS observations have detected $\gamma$-rays from SN1006 with energies of about 10 TeV, requiring at least 70 TeV protons if the dominant process is through the hadronic channel. If the emission is of leptonic origin, the electron energy should be around 50 TeV for the inverse compton process of the microwave background, which has a characteristic photon energy of $\epsilon_1=4 \gamma^2 \epsilon_0$ \citep[e.g.][]{1970BlumenthalGould}, with $\gamma$ the lorentz factor of the relativistic electron, and $\epsilon_0$ the initial photon energy. With our model parameters, from Eq.~\ref{eq:emax} we get a value of $\sim 60$~TeV for the maximum cosmic ray energy, which is still in the right range, but that is assuming that the NRH instability still operates. If and how other instabilities will be efficient enough to reach similar cosmic ray energies is an open question. It is possible that currently observed cosmic rays were accelerated earlier on, and whether the spectrum has started to roll over or not at these energies also makes a large difference. Alternative solutions could be a slightly higher density, lower background magnetic field strength, or a higher value for the initial fluctuations, all of which will help to continue the NRH for a longer period of time and keep $E_{max}$ higher. All in all, we cannot be quite certain that the NRH instability is still operating in SN1006, and it may potentially be a good candidate to look at what happens beyond the NRH instability. Modelling of the spectrum based on assumptions for the electron and proton population of cosmic rays has been done by various authors \citep[e.g.][]{2012Berezhkoetal} and we refer to those works to see how the cosmic ray spectrum translates to the photon spectrum in more detail.

Our model, although it does not cover the downstream values, gives a reasonable indication of which SNRs are efficient in magnetic field amplification and which are less so. 
For the SNRs exploding in a CSM the assumption of a homogeneous, parallel, field of $5\;\mu$G is obviously overly simplified. The farther the shock wave propagates, the more perpendicular the magnetic field it encounters will be as the shock propagates further into the Parker spiral generated by the wind of the rotating progenitor star. Its zeroth order field strength is therefore also likely to be a function of radius.


During the nonlinear stage, there is also the possibility that other instabilities will come into play. Given there is now a cosmic ray precursor, the acoustic instability as described by \citet{2012DruryDownes, 2009MalkovDiamond}, or a filamentation instability such as described by \citet{2011RevilleBell}, could start to act on the conditions created by the NRH instability. As these instabilities are longer wavelength in nature, they may aid in confining cosmic rays up to higher energies. When the shock crosses this region of density and magnetic field fluctuations it compresses the perpendicular component of the field, and additional amplification may occur downstream \citep{2011Beresnyak,2012Guoetal} as a result of the (NRH or otherwise induced) density fluctuations.

\section{Galactic pevatrons and cosmic ray spectrum}
\label{sec:cumulative}

We have seen in the previous section that the highest energies for cosmic rays are reached in early phases of the SNR evolution, and especially in the core-collapse SNe in a dense RSG wind, which are representative for the early stages of most type II SNe. However, with our assumed model parameters we get to about a PeV but not to much beyond, and only for SNRs younger than a few decades. Potentially, if a proper description for the magnetic field around a massive star is taken into account, this may shift the numbers. 

There are a couple of ways to increase the cosmic ray energy compared with the current analysis. Firstly, some change in $E_{max}$ may be gained by adjusting the explosion parameters: mass and energy of the ejecta. Pushing the mass to an extreme low and the energy to a high will increase the maximum cosmic ray energy some -- by virtue of the higher shock velocity -- depending on the density of the environment. 
Secondly, the energy of the cosmic rays may be increased by the inclusion of higher Z elements, as was also argued by \citet{2010Ptuskinetal}. For example, if the wind were dominated by helium rather than protons, the energy would increase twofold. Observations of the cosmic rays seem to indicate that the mean mass of the cosmic rays seems to go up between several 100s of TeV and 10 PeV \citep{2012KampertUnger,2012Icecube}, which is very interesting in light of these results. The systematic uncertainties are quite large and the interpretation model dependent, which allows for a wide range of energies at which the composition might change. However, both of these methods to increase the cosmic ray energy are mostly applicable to the case we describe for a WR wind -- in a tenuous environment the higher shock velocity survives longer, and the dominance of helium versus hydrogen may be expected. Because of the low density resulting from the WR wind, the cosmic ray energies are very low to start with, and even with the increases that will result from the proposed changes, we do not expect they will get to PeV energies.

The confinement from the NRH instability may continue as long as the gyroradius of the highest-energy cosmic rays is larger than the wavelength of the fastest growing mode. It turns out that, in our models, with the simplified assumption of having a radial magnetic field with a magnitude of $5\;\mu$G, this condition ceases to be met after around $\sim 1000$~yrs, depending on the exact parameters. For a wind-shaped CSM, if the magnetic field strength continues to decrease, this could be at a later time. However, by then the perpendicular component of the magnetic field will be so dominant compared to the parallel component that we are in a different regime altogether, which requires a new line of investigation. If SNRs continue to accelerate CRs beyond this point, the resonant and/or long-wavelength instabilities \citep[e.g.][]{2011SchureBell} may dominate. 

\begin{figure}
  \centering
\includegraphics[width=0.49\textwidth]{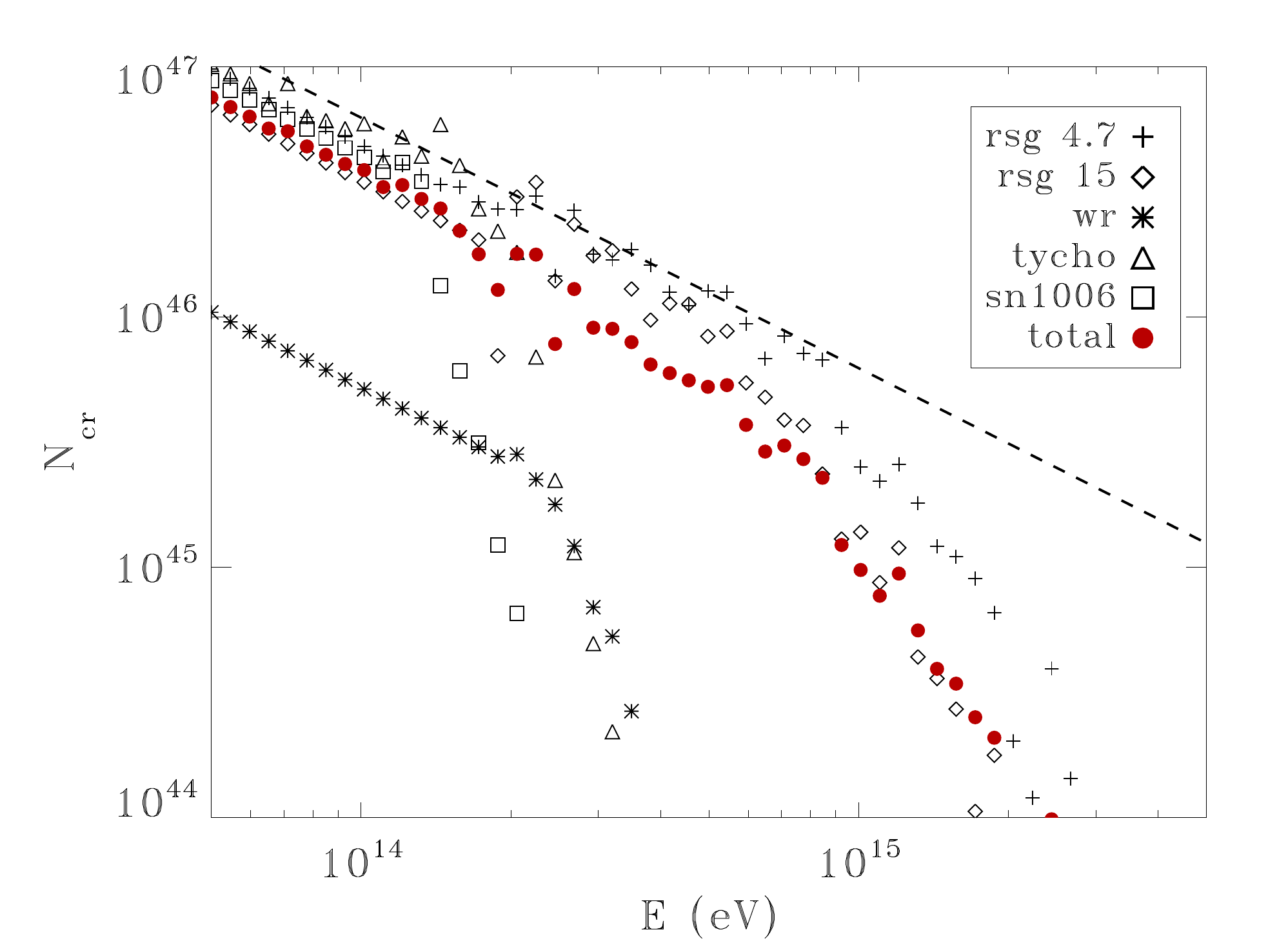}
 \caption[ ] {Cosmic ray spectrum of the escaped particles, extrapolated as a $\propto E^{-2}$ power law to lower energies from the energy where the NRH stops working (compare to the energy where the curve in Fig.~\ref{fig:Emaxall} drops). The cumulative spectrum is shown with red dots, with fractions of 1/2 RSG, 1/6 WR, 1/6 SN1006-like, and 1/6 Tycho-like SNRs. The dashed line indicates the slope assumed for the spectrum at the shock, following a powerlaw of $dN/dE\propto E^{-2}$. 
     \label{fig:cumulative}}
      \end{figure}

We can determine the resulting spectrum from escaping cosmic rays by adding the cosmic rays that escape at any given time to the escape spectrum, and calculate the resulting spectrum over the evolution of the SNR (up to the age to which we can track it, which gives a low-energy bound to our spectrum). We do this for each of the types of SNRs that we consider separately, and show the results in Fig.~\ref{fig:cumulative}. The number of escaping cosmic rays is essentially determined by the escape current as $N(E)=j/(u_s q)$, integrated over the surface area of the entire remnant, while the energy $E_{max}$ of the escaping cosmic rays is determined by Eq.~\ref{eq:emax} for a spectrum at the shock that follows a slope $\propto E^{-2}$. A monochromatic energy spectrum at this energy is added to the energy bin with $E<E_{max}<E+dE$, to contribute to the escape spectrum. We extend the spectrum to energies below the energy reached at the time our NRH instability stops operating, following a slope of $d N_{cr}/dE \propto E^{-2}$, which is the slope expected based on the dynamics once the Sedov phase is reached \citep[e.g.][]{2013Belletal}. The non-smoothness of the spectrum is a result of the limitations of tracking the shock precisely and the discretisation, especially at early times. At the high-energy end the spectrum is significantly steeper than $E^{-2}$, as expected for the free expansion phase. \citet{2005PtuskinZirakashvili} use the Chevalier-Nadyozhin solution to predict slopes during the free expansion phase, values of which would amount to  $E^{-4}$ and $E^{-5}$ for our used CSM and ISM parameters respectively, in the free expansion phase. Especially for the CSM cases, the gradual conversion to a $E^{-2}$ power law at lower energies can be seen to emerge.
By combining all of the generated spectrum and adding the escaping cosmic rays, we can get a rough estimate of the source spectrum in our Galaxy, plotted in Fig.~\ref{fig:cumulative} indicated by the red dots. We have used the following rates: $1/6$ Tycho-like events, $1/6$ SN 1006-like Type Ia SNe,  $1/6$ WR-wind CSM core-collapse events, and $3/6$ RSG wind CSM. These numbers are roughly in correspondence with some recent SN rates surveys \citep{2009Smarttetal, 2011Leamanetal}. 
The slope of the total spectrum is slightly steeper than the source spectrum, indicated by the dashed line, and will be smoother if a greater variation of explosion and ISM/CSM parameters will be taken into account.

From our model we can calculate the energy carried away by escaping cosmic rays for the different systems (up to the time the NRH instability stops operating, indicated by the steep drop in Fig.~\ref{fig:Emaxall}). The total energy pumped into cosmic rays up to this time can estimated by extrapolating the energy spectrum to the rest mass energy, assuming that the powerlaw extends to lower energies with a slope of $E^{-2}$. The energy in respectively the escape spectrum, and the `total' spectrum varies between $1.8 \times 10^{49}$~erg (total: $\sim 10^{50}$~erg) for a Tycho-like event in the first 900~yrs, $1.0 \times 10^{49}$~erg (total: $8.7 \times 10^{49}$~erg) for a SN 1006-like system in the first 500 years, $3.5 \times 10^{49}$~erg (total: $1.2 \times 10^{50}$~erg) in 1000 years for a dense RSG wind, and just $1.8 \times 10^{48}$~erg (total: $\sim 10^{49}$~erg) for a system in a WR wind in 200 years, after which the circumstellar medium becomes more complex. From these numbers we can deduce that during the operation of the NRH instability, on average only $\sim1$\% of the total kinetic energy has been lost in the form of escaping cosmic rays, but a total of on average about 5.5\% of the kinetic energy of the shock has been transferred to cosmic rays up to this point (amounting to an average of $\sim 9 \times 10^{49}$~erg). More energy will be transferred to CR during the subsequent Sedov phase of the SNR expansion.

It seems that of these type of systems, a core-collapse in a RSG environment not only accelerates to the highest energies, but also produces the majority of the cosmic rays because of the higher explosion energy and the high circumstellar density. Our model predicts that in these systems, for our chosen parameters, about $6\%$ of the kinetic energy has been transferred to cosmic rays in the first approximately 1000 yrs during which the NRH instability is expected to operate. This is in the case of the very slow (and thus high density) wind.

The total energy required to replenish the Galactic population of cosmic rays has been estimated to be on the order of $3 \times 10^{40}$~erg~s$^{-1}$ \citep[e.g.][]{1987BlandfordEichler,1984Lund}. For our model parameters, we are able to reach this power even if the supernova rate is only about two per century. However, estimates of the required CR energy input to the Galaxy are very uncertain, and estimates of the required supernova rate are correspondingly uncertain. 

The highest cosmic ray energies are reached in the first couple of hundred years in a dense wind environment, after which it rapidly drops.
For the known Galactic SNRs, Cassiopeia A is the best candidate to have been a `Pevatron', albeit only in its early years and not currently. Another candidate is potentially G1.9+0.3, but its circumstellar density is expected to be low \citep{2011Carltonetal}, which likely prevents it from acceleration cosmic rays to PeV energies, and, indeed, from being a significant source of the highest energy galactic cosmic rays.

\section{Amplification and cosmic ray energy in SNRs}
\label{sec:SNRmodels}

Finally, we will look at how the NRH instability manifests itself using the values for the escape current found in the calculations for self-consistent cosmic ray confinement and magnetic field growth in Sect.~\ref{sec:Emax}.

\begin{figure}
  \centering
\includegraphics[width=0.45\textwidth]{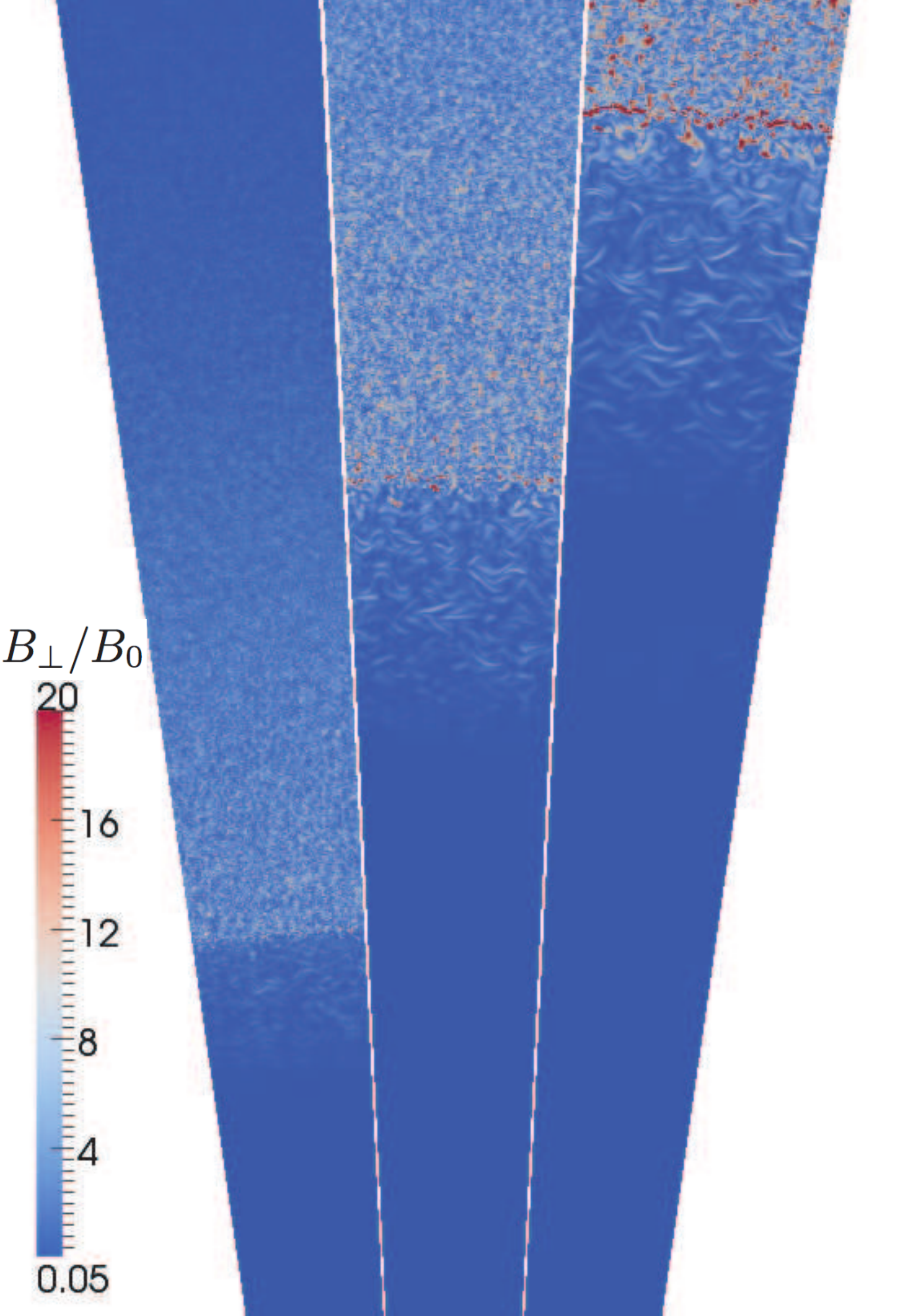} 
\includegraphics[width=0.45\textwidth]{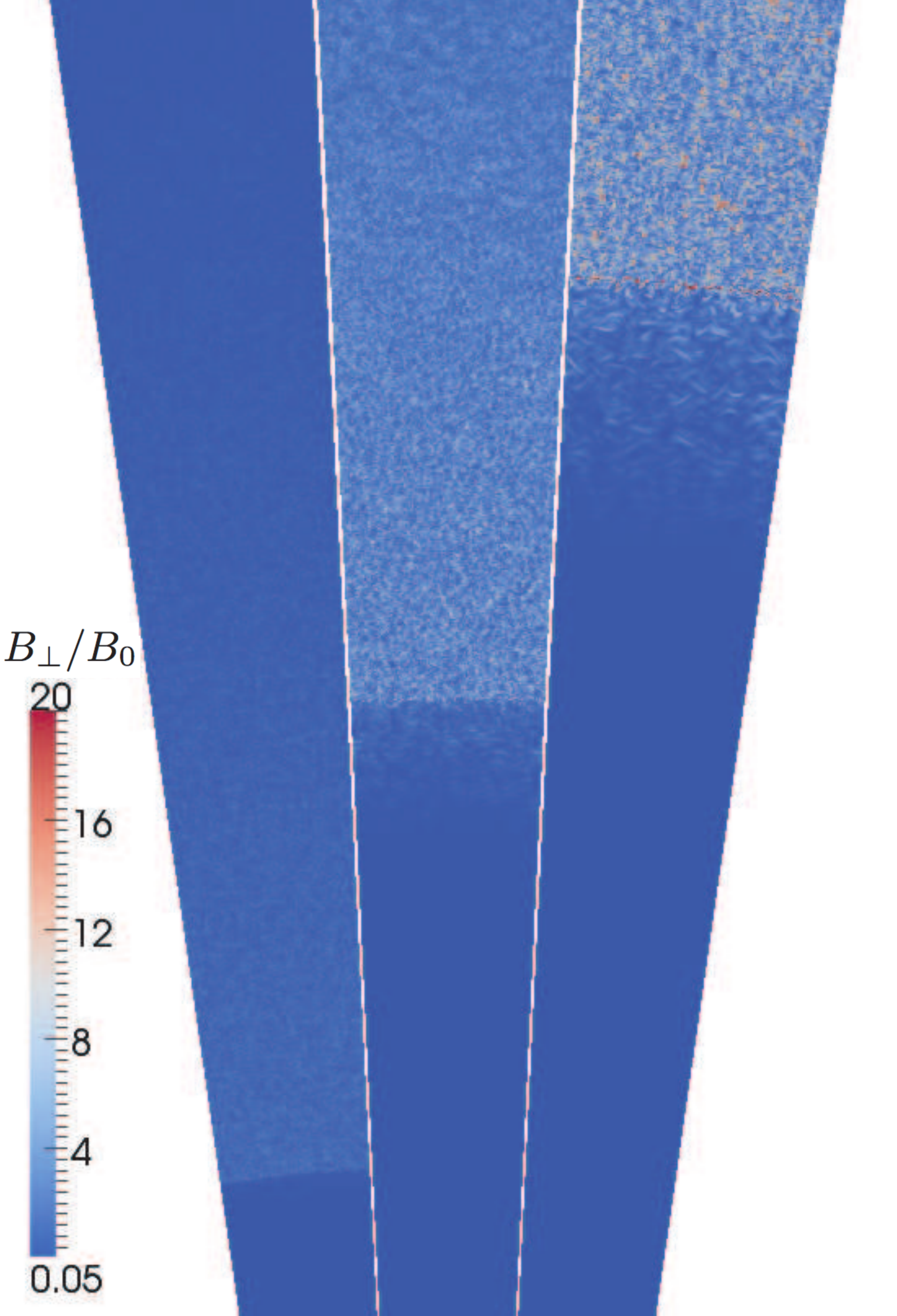}
 \caption[ ] {$B_\perp/B_0$ at different times for SNR evolving in a ISM (upper panel) and CSM (lower panel), for $t \approx 200$, $300$, and $400$~yr. The shock radius at those times is $2.1$, $2.65$, $3.1$~pc (ISM), and $1.85$, $2.4$, $2.9$~pc (CSM).
    \label{fig:largescale}}
  \end{figure}

\begin{figure}
  \centering
\includegraphics[width=0.45\textwidth]{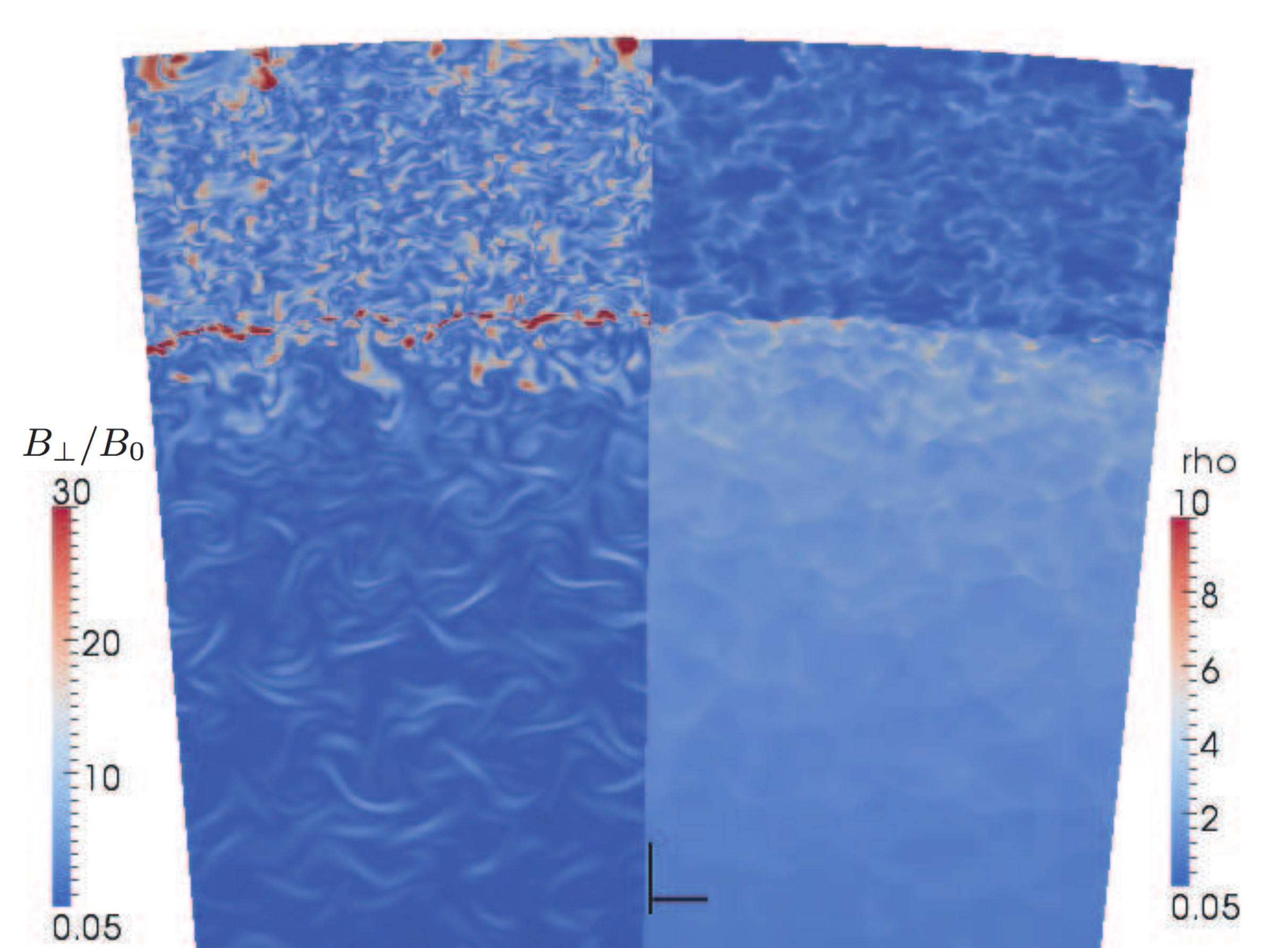} 
\includegraphics[width=0.45\textwidth]{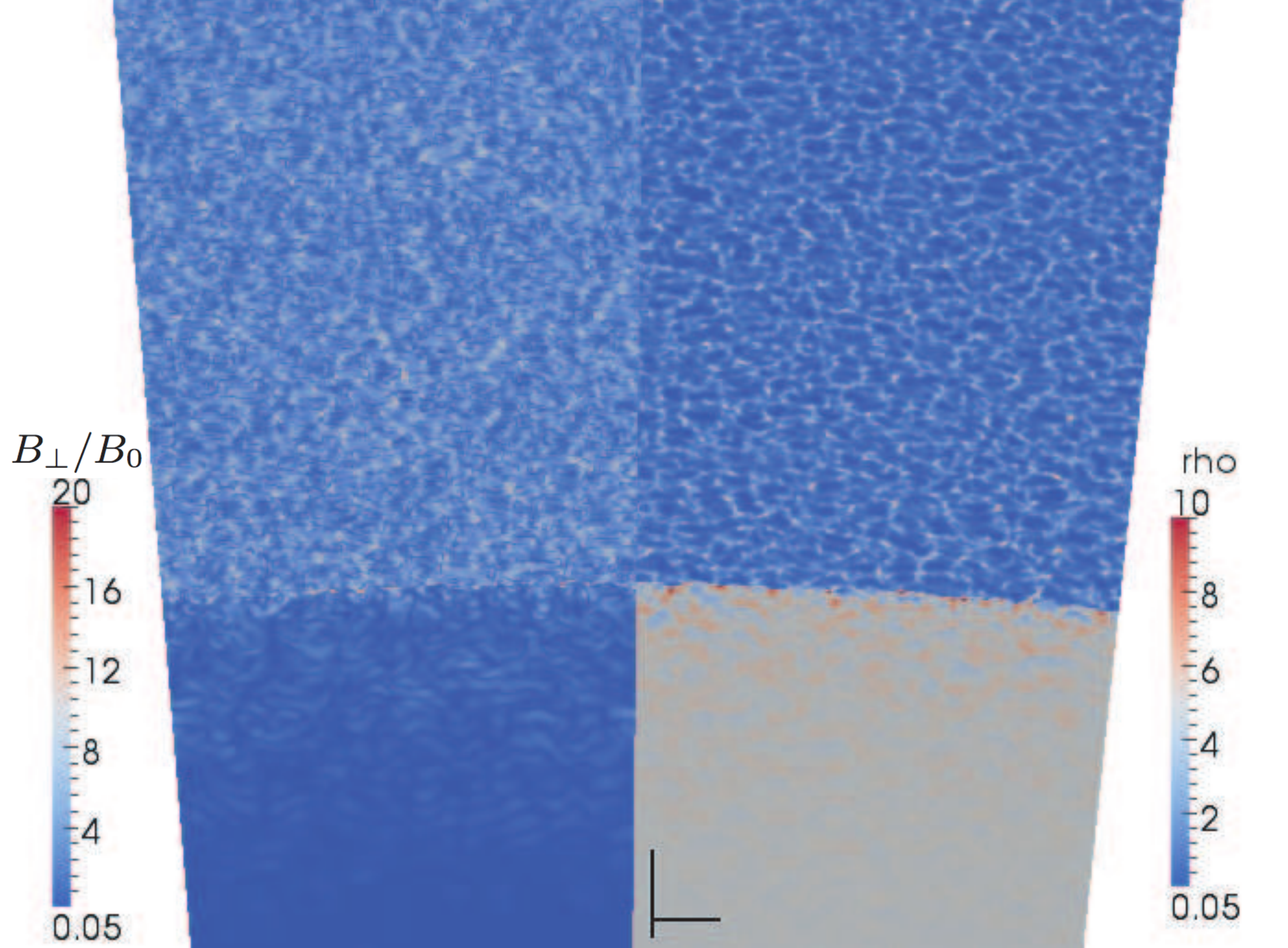}
 \caption[ ] {$B_\perp/B_0$ and density are shown for the SNR shock region of a Tycho-like SNR (top), at a time of $\sim 400$~yr, and Cas A like SNR at a time of 300~yr. The axes indicate a scale of $10^{17}$~cm.
    \label{fig:closeuptycho}}
  \end{figure}

In order to see the large scale evolution, we have modelled the evolution of a SNR using the AMRVAC code discussed in Sect.~\ref{sec:model}. We set the simulation up in 2.5D spherical $r$, $\theta$ geometry, with symmetry in the $\phi$ direction, on a grid covering an angle of $5.4^\circ$ around the equator and a radial extent of $\sim 3.2$~pc. We use 10440 cells in the radial direction, and 348 in the angular direction, resulting in a radial resolution of $\Delta r \approx 9.6 \times 10^{14}$~cm. A CR current is imposed upstream of the shock, according to the 1D model described in Sect.~\ref{sec:Emax}. We have run simulations with the same Tycho- and RSG-like parameters as discussed in the previous sections. Because of numerical reasons, we initialise the ejecta over a larger initial radius ($1.5 \times 10^{18}$~cm) with a lower maximum velocity ($15,000$~km~s$^{-1}$), in order to obtain a solution for the shock position that is as close as possible to that in our 1D runs. Apart from that, the initialisation is as described in Sect~\ref{sec:Emax}. The inner boundary is set at $0.5 \times 10^{18}$~cm, to avoid inifinitesimal time step near the origin.  This means we have a non-zero velocity at the inner grid, which causes non-homologous expansion of the inner ejecta. However, this does not affect the region beyond the contact discontinuity much and certainly does not reach to the forward shock. We use Eq.~\ref{eq:j_gt} to calculate the current, where we again choose $\gamma \tau=5$, which is then used as input for the Lorentz force in the momentum equation. The current and $B_0$ are set up in the radial direction, and a random magnetic field is added to that, with values of again 5\% of the background magnetic field. The result is shown in Fig.~\ref{fig:largescale}, where we show $B_\perp/B_0$ for a part of the grid at times 200, 300, and 400 yrs (left to right). The upper panel shows the Tycho-like SNR, and the lower panel the RSG (or Cas A)-like SNR. 

\begin{figure*}
  \centering
\includegraphics[width=0.48\textwidth]{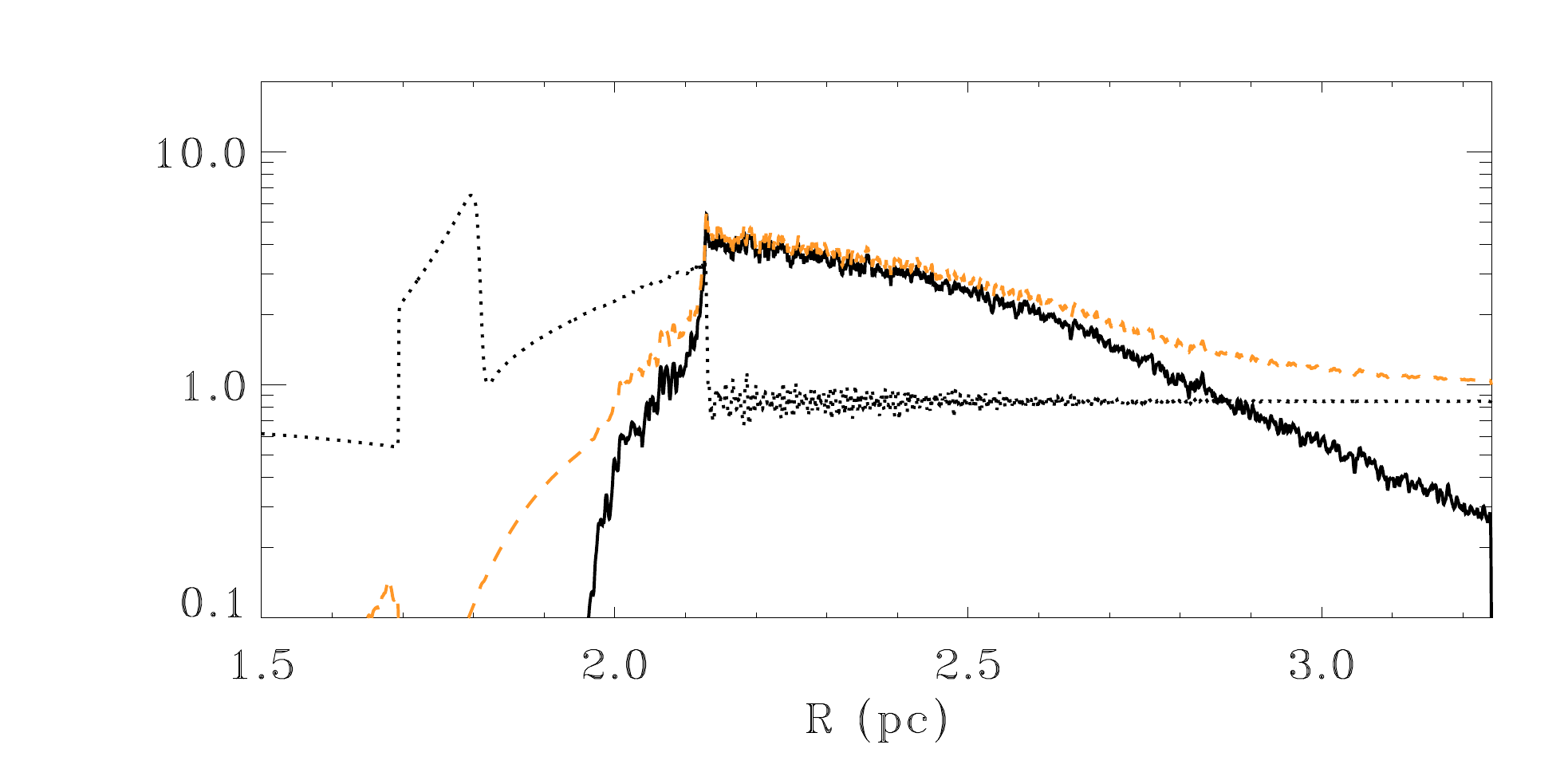}
\includegraphics[width=0.48\textwidth]{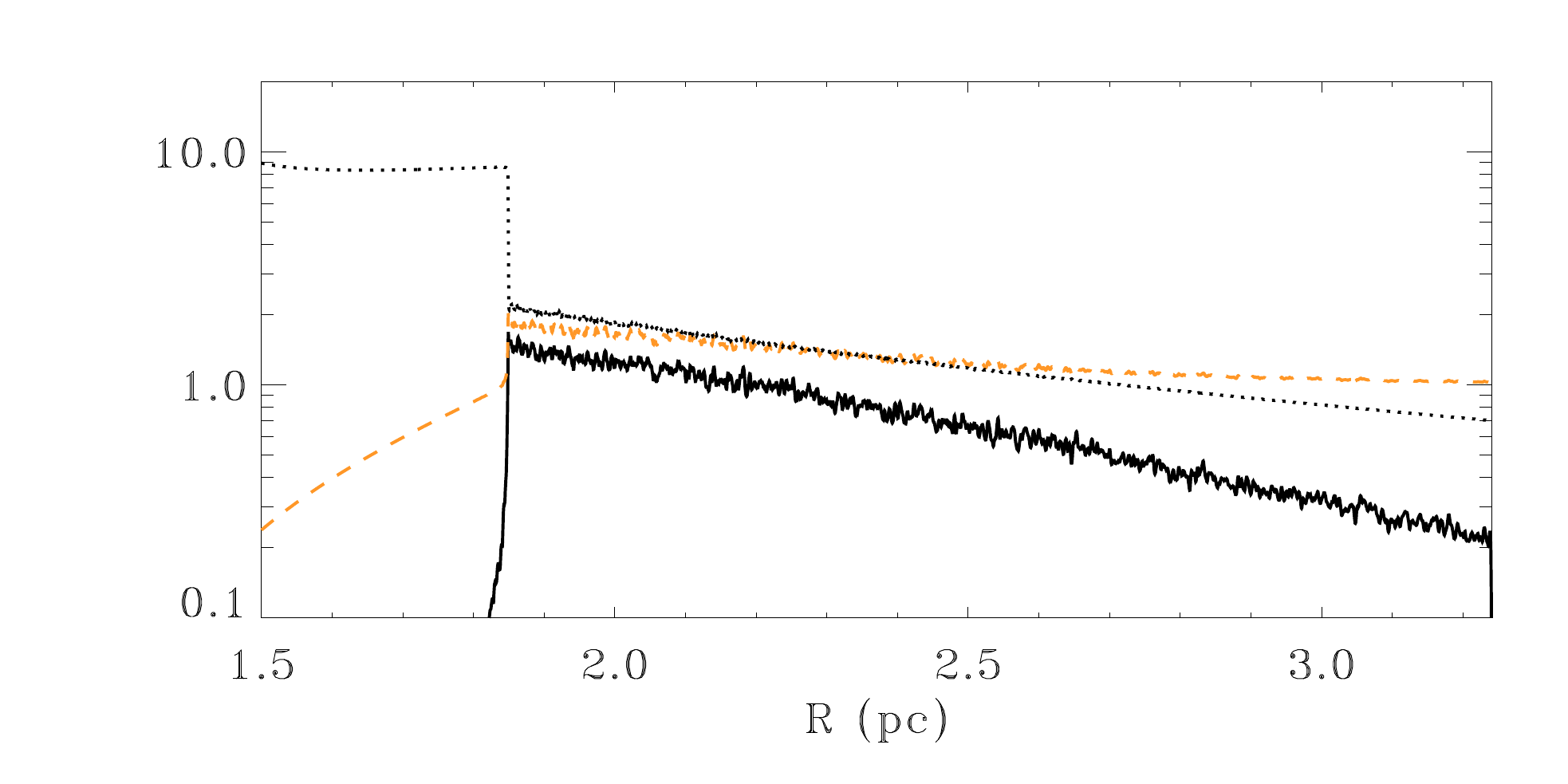}
\includegraphics[width=0.48\textwidth]{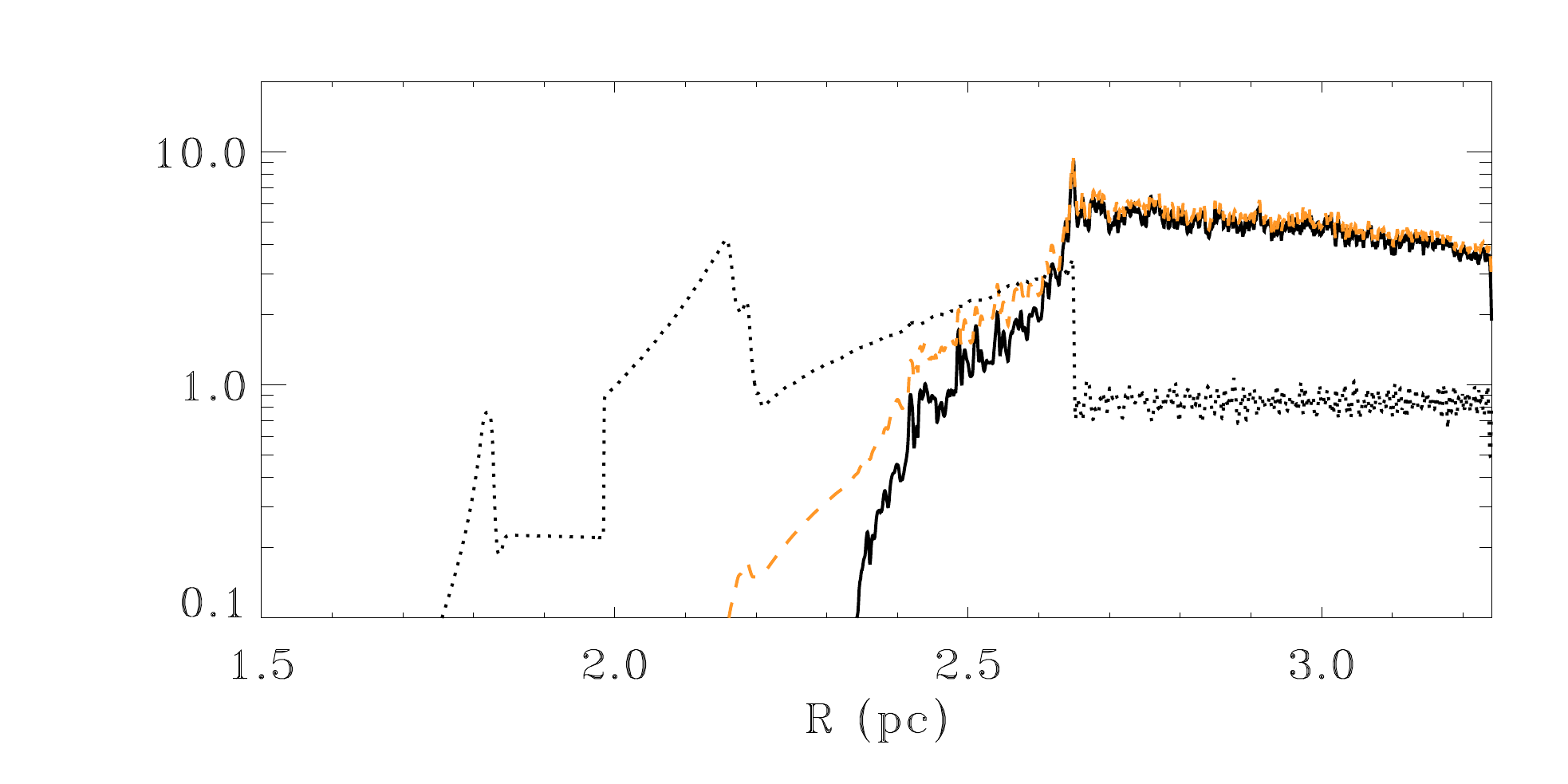}
\includegraphics[width=0.48\textwidth]{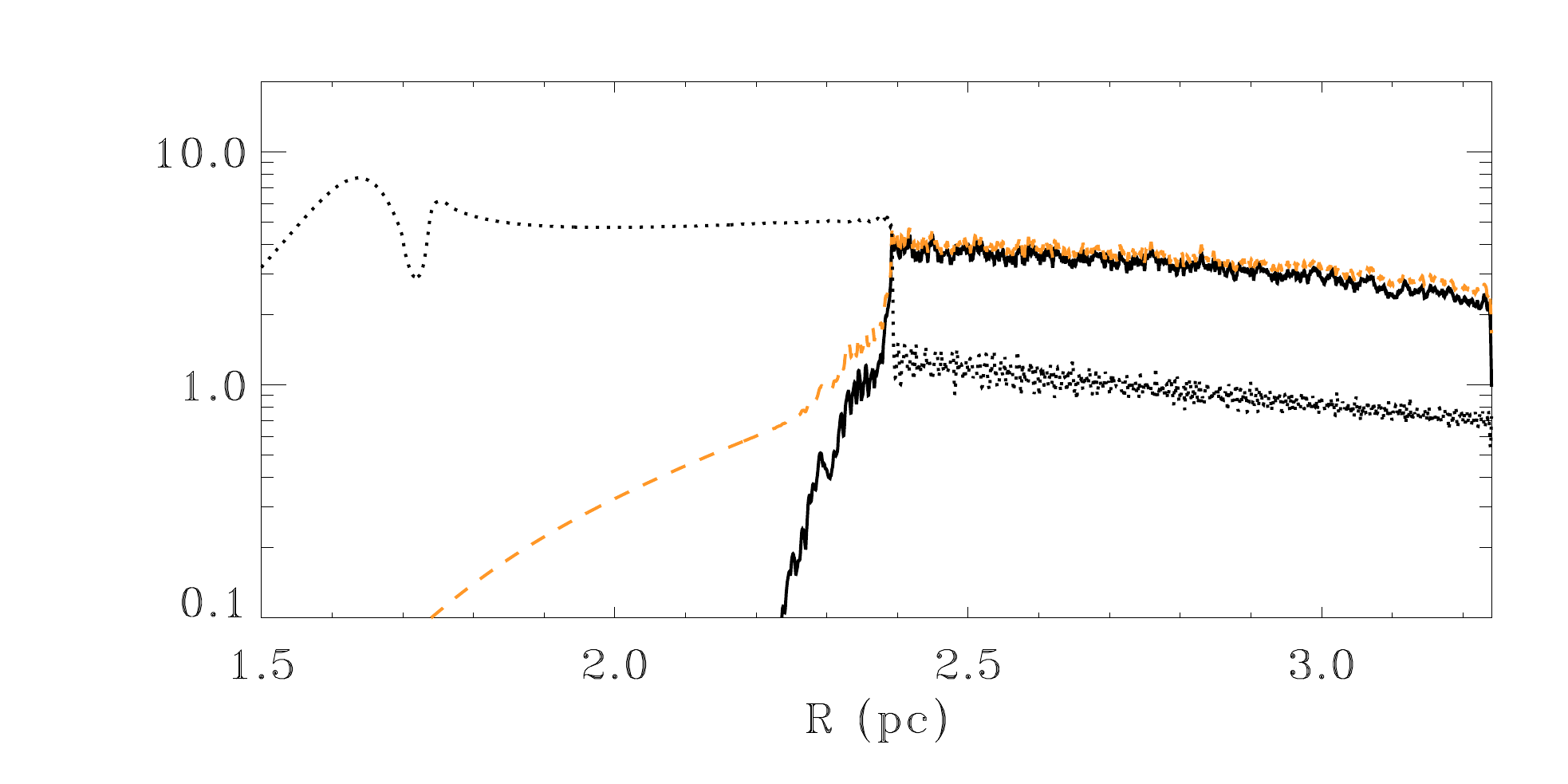}
 \caption[ ] {$\langle B_\perp\rangle/B_0$ (solid), $B_{rms}/B_0$ (yellow dashed), and density (dotted) are shown as a function of radius, averaged over the angular extent of the simulation, for a Tycho-like (left) and Cas A like (right) SNR at $t=200$ (upper) and $t=300$~yr (lower panels).
    \label{fig:averagebTycho}}
  \end{figure*}

An angular average as a function of radius of the density and of $B_\perp$ and $B_{rms}$ is shown in Fig.~\ref{fig:averagebTycho} for both types of SNR, at times $t=200$ and $t=300$. At $t=200$, in both cases the gradual growth of the upstream magnetic field can be seen, while at a later time the average field strength has mostly saturated. Downstream, the magnetic field decays rapidly. The jump of $B_\perp$ at the shock is better viewed in the 2D plots of Figs.~\ref{fig:largescale} and \ref{fig:closeuptycho} because the average over angle, combined with the shock corrugation, smooths out the jump in $\langle B_\perp \rangle$. 

Theoretically, at any point in time, the growth just upstream of the shock should be such that $\gamma \tau$ reaches a maximum value of $5$, resulting in a value of $B_{\perp,max}/B_0\approx 7.5$ for exponential growth. At early times the growth seems to progress more slowly, whereas later on the amplification is stronger. We find that this is consistent with what we derive using our numerical iteration for solving for $\gamma \tau$ in Sect.~\ref{sec:Emax} (see Fig.~\ref{fig:Emaxall}), using the shock position in the simulation. Because in 2D we cannot track the shock at early times as precisely as we did in our 1D simulations, because we initialise the ejecta over a larger radius, we find that at the very early times $E_{max}$ from Eq.~\ref{eq:j_gt} is higher than we find numerically, resulting in a current that is lower than needed to reach the required levels of $\gamma \tau$. Later on, the reverse happens, and the analytical results yields a current that is slightly larger than that needed to reach $\gamma \tau = 5$. Also, in the ISM environment, growth is relatively more rapid. In Fig.~\ref{fig:Emaxall} the analytical curve is further below the numerical outcome because $\ln(E_{max}/E_{rest}) < 14$, resulting in underpredicting $E_{max}$ and thus overpredicting $j$, resulting in higher levels of magnetic field amplification. As said before, small changes in the current induce large changes in the amplification, and we consider the outcome consistent with expectations. An additional factor that may be of influence in generating higher levels of $B_\perp/B_0$ is that, although the initial average strength of the seed magnetic fluctuations is on the 5\% level, the peak levels of the initial fluctuations are substantially higher than that.
We should also consider that in the simulations we do not distinguish between the scale of the maximum growing mode and the scale of the seed fluctuations, whereas in the simulations those scales change over location and time.

A close up of both SNRs, which also depicts the density, is shown in Fig.~\ref{fig:closeuptycho}. ``Tycho'' is plotted for a time $400$~yr, and ``Cas A'' is plotted at $t=300$~yr (for even better close-ups we refer to the corresponding $B_\perp/B_0$ frame in Fig.~\ref{fig:fixedcurrent}, i.e. the bottom three frames). The axes indicate a scale of $10^{17}$~cm. Values of $B_\perp/B_0$ of $10-25$ have been reached, corresponding to upstream magnetic field strengths of $50-125\;\mu$G. For these amplification levels, clear density contrast can be seen in the upstream, and even after shock passage. Such density perturbations may induce additional instabilities, both in the downstream \citep{2007GiacaloneJokipii, 2008ZirakashviliPtuskin,2012Guoetal}, and in the precursor \citep{2011Beresnyak, 2012DruryDownes}. All these simulations, however, are done in slab geometry and do not account for expansion of the plasma. In our simulations we do see a stronger magnetic field at the shock, compared to the upstream, but whether this is different from just compression is unsure -- if it happens it may be at a smaller scale than we are interested in. We mainly see that further downstream the magnetic field strength decays, possibly a combination of the compressed magnetic field relaxing now that there is no $j \times B$ force confining the high-pressure regions of dense plasma and amplified field. Additionally, the post-shock reduction in the magnetic field may be increased by expansion in spherical geometry and possibly also by numerical diffusion caused by rapid downstream advection accross the computational grid.

Observations of non-thermal X-ray emission \citep{2004Hwangetal} of the shock front of Cas A show that there is a fair bit of structure at the shock. The thin filaments that are observed in X-ray wavelengths correspond to length scales of the same order of magnitude as what we see in Fig.~\ref{fig:closeuptycho}, around $2.5 \times 10^{17}$~cm \citep{2009PatnaudeFesen}. Although our simulations do not show the same level of structure as the observations do, a small adjustment to the current density could increase the level of structure substantially. 

In our simulations more structure is apparent in the Tycho case, whereas observations for Tycho show a smoother shock front for the most part. This could mean that for Tycho the current density in the simulation should be slightly lower than the value that we used, or other parameters could be changed to achieve the same result. It is worth noting that in this approach, the current is only dependent on the values for $\gamma \tau$, $R_s$, $u_s$, and $\rho$. Only the corresponding value for $E_{max}$ depends additionally on the efficiency $\chi$. Observations of cosmic ray energies can therefore constrain $\chi$, whereas observations of the upstream structure and magnetic field value can teach us something about $\gamma \tau$ and the current density. It is possible that a substantial level of structure contains a signature of the present activity of the NRH instability and could be used to constrain the associated current density. However, spatial structure and time variation might be caused by the shock propagating into an already disturbed medium, for example as indicated by the presence of the quasi-stationary flocculi being overtaken by the shock in Cas A.

The magnetic field values that we reach in our simulations (Fig.~\ref{fig:closeuptycho}) and that we have derived from the equations are, in the case of Tycho and Cas A, short of the saturation value, which would amount to $\sim 400$ and $\sim 980\;\mu$G respectively ($\sqrt{11}$ times the upstream saturation value given by Eq.~\ref{eq:Bsat}). This is unsurprising since amplification of $B_\perp$ by a factor $e^{\gamma \tau}$ with $\gamma \tau=5$ and further increase by $\sqrt{11}$ through compression at the shock would only increase the field to $\sim 120\;\mu$G. So long as this is the case, the magnetic energy density is likely to scale with $\rho u_s^2$. Only when saturation is reached, such as may be the case in SN 1006, we expect $B^2$ to scale with $\rho u_s^3$.

\section{Discussion}
\label{sec:discussion}

We have attempted to find a combination for the current and the maximum cosmic ray energy that simultaneously allows for confinement of cosmic rays and growth of the magnetic field. We expect the non-resonant hybrid instability to be the most efficient at amplifying the magnetic field in young SNRs, up to an age of $\sim 1000$~yrs. Based on the maximum growth rate of this instability, and of the required growth of the magnetic field, the value of the cosmic ray escape current can be derived that is consistent with such growth of the upstream magnetic field. This, combined with an efficiency parameter $\chi$, gives us the maximum energy of cosmic rays at any point in time that is consistent with such a current.

In historical SNRs such as Tycho and Cas A, there is strong evidence of an amplified magnetic field. Using these to help constrain our input for the minimum value of $\gamma \tau$ and the corresponding required minimum value of the current density, an upper limit for $E_{max}$ can be derived. We find that values are consistent with these SNRs currently being sources of cosmic rays to energies of $\sim 100$ and $300$ TeV respectively, depending on the exact choice of parameters. These values are consistent with available observations, and we hope that the upcoming Cherenkov Telescope Array \citep{2013CTA} will be able to actually determine at which energy the gamma-ray spectrum rolls over, giving us a much better handle on the expected cosmic ray energy.

SNRs that evolve in a high-density environment are more effective in accelerating cosmic rays to high energies. In addition to having high initial densities, core-collapse SNe gain an extra factor 2 (see Eq.~\ref{eq:emax}) for the cosmic ray energy because of their stellar-wind environment. This results from the fact that the current is proportional to the density at the shock front, whereas the upstream growth rate is inversely proportional to the square root thereof, resulting in an overall net gain of a factor two when it comes to the maximum energy of cosmic rays.  

The historical remnant of SN1006 is of Type Ia and, according to our model, has reached a point where the NRH instability may no longer be operating. The value of the measured magnetic field is much lower than that for the other historical remnants, which also is an indication that cosmic ray acceleration is not very efficient. Parameters may be slightly altered, such that SN1006 still fits within our model, or alternative instabilities should be explored that keep the cosmic rays near the shock front. 

When it comes to the total population of cosmic rays arriving from our analysis, we find that, in the first 1000 years of the lifetime of a SNR, with our assumed model parameters, on average about 5.5\% of the kinetic energy has been converted into cosmic rays (in the best possible accelerator: the RSG wind, this value is 6\%). Even with a supernova rate of about two per century, this is potentially enough to replenish the Galactic population, and there is room for the efficiency to be below our optimistically assumed value. Our results are consistent with SNRs being able to be responsible for the dominant component of Galactic cosmic rays. 

It is unlikely that at a later times the maximum energy of cosmic rays will be higher than that resulting from this analysis. The population of PeV cosmic rays is most likely to come from young SNRs of the SNe that explode in a dense wind medium. In our model, we find that energies of a few PeV are reached. Given the high initial density, and therefore potential number of cosmic rays, the resulting escape spectrum still follows a powerlaw, that is only mildly steeper than the source spectrum.

In conclusion, the combination of diffusive shock acceleration with the non-resonant hybrid instability provides a fairly self-consistent picture for the origin of CRs up to PeV. Given sufficient energy transfer in the first 1000 yrs, the process of DSA proceeds rapidly enough to accelerate the high-energy end of the population of Galactic cosmic rays. Special circumstances are required to get beyond energies of a few PeV. Escape is an essential requirement for acceleration to high energies, as the instability and confinement of cosmic rays to the region are mutually dependent. The exact diffusion processes and other instabilities that will start to dominate for ageing SNRs will be important in addressing the spectrum of Galactic cosmic rays below $\sim 100$~TeV. When it comes to PeV cosmic rays though, the NRH instability makes young SNRs, in combination with a dense RSG environment, the most likely candidates for their origin.

 \section*{Acknowledgements}
The research leading to these results has received funding
from grant number ST/H001948/1
made by the UK Science Technology and Facilities Council and from the European Research Council under the European
Community's Seventh Framework Programme (FP7/2007-
2013)/ERC grant agreement no. 247039.

\bibliography{../adssample}

\label{lastpage}
\end{document}